\documentclass[a4paper,12pt]{article}

\usepackage{jheppub} 

\usepackage[T1]{fontenc} 

\usepackage{float, array, xspace, amscd, amsthm}
\usepackage{fancyhdr}
\usepackage{longtable}

\newcommand{\dd}{{\rm d}}

\usepackage[usenames,dvipsnames,svgnames,table,x11names]{xcolor}
\definecolor{MagentaXD}{RGB}{204, 48, 152}
\definecolor{MagentaXDdetail}{RGB}{150, 79, 126}
\definecolor{GreenMAF}{RGB}{28, 112, 46}
\definecolor{GreenMAFdetail}{RGB}{80, 117, 88}
\definecolor{detail}{RGB}{110,110,110}

\newcommand{\be}{\begin{equation}}
\newcommand{\ee}{\end{equation}}
\def\bea#1\eea{\begin{align}#1\end{align}}
\newcommand{\tr}{{\rm tr}}

\newif\ifcomments
\commentstrue

\newif\ifdetails
\detailstrue

\usepackage{upgreek}
\usepackage[normalem]{ulem}
\usepackage{mathtools}
\usepackage{datetime}
\usepackage{cancel}

\settimeformat{ampmtime}

\title{
$\mathcal{G}$-structure symmetries and anomalies in $(1,0)$ non-linear $\sigma$-models
}

\author[a]{Xenia de la Ossa,}
\author[a]{Marc-Antoine Fiset}

\affiliation[a]{Mathematical Institute, University of Oxford\\Andrew Wiles Building, Woodstock Road\\Oxford OX2 6GG, UK }

\emailAdd{delaossa@maths.ox.ac.uk, fiset@maths.ox.ac.uk}

\abstract{
A new symmetry of $(1,0)$ supersymmetric non-linear $\sigma$-models in two dimensions with Fermi and mass sectors is introduced. It is a generalisation of the so-called special holonomy $W$-symmetry of Howe and Papadopoulos associated with structure group reductions of the target space $\mathcal{M}$. Our symmetry allows in particular non-trivial flux and instanton-like connections on vector bundles over $\mathcal{M}$.
We also investigate potential anomalies and show that cohomologically non-trivial terms in the  quantum effective action are invariant under a corrected version of our symmetry. Consistency with heterotic supergravity at first order in $\alpha'$ is manifest and discussed.
}

\begin{document} 

\maketitle
\flushbottom

\section{Introduction}

Bootstrap approaches using symmetries are a mainstay of modern research, providing unparalleled glimpses into the exact nature of quantum field theories. Meanwhile, the Lagrangian-based formalism, when it exists, remains arguably the most concrete handle on the theory. It is of interest, when possible, to compare these frameworks.

In two dimensions, a further reason to study this connection\footnote{Similar considerations can be found in \cite{Wendland:2015rla}.}
stems from the intimate relationship of supersymmetric Lagrangians, or non-linear $\sigma$-models, with the geometry of their space $\mathcal{M}$ of field configurations. This may inform us on string dualities generalising mirror symmetry, amongst other applications; see e.g.\ \cite{Blumenhagen:1996vu,Shatashvili:1994zw,Acharya:1997rh,Gaberdiel:2004vx, Melnikov:2012hk,Braun:2017ryx,Braun:2017csz,Fiset:2018huv}.
A particularly interesting case is when the target space of the $\sigma$-model is a manifold with reduced $\mathcal{G}$-structure. Related to these, there exist, in the conformal case, exact field theory descriptions based on chiral symmetry $W$-algebras. The most well-known example is when the target space is a complex manifold, corresponding to the $N=2$ Virasoro algebra \cite{Zumino:1979et}. Other more intricate algebras, in particular related to exceptional structure groups \cite{Shatashvili:1994zw,Figueroa-OFarrill:1996tnk}, have also appeared in this context \cite{AlvarezGaume:1981hm,Odake:1988bh}.

It is natural to ask for a characterisation of the precise $\mathcal{G}$-structure target manifolds $\mathcal{M}$ related to a given $W$-algebra, at least to leading orders in $\sigma$-model perturbation theory. At the classical level, a long-established result due to Howe and Papadopoulos \cite{Howe:1991ic,Howe:1991vs,Howe:1991im} sheds light on this question in the context of massless $(1,0)$ non-linear models with generic target space metric and B-field \cite{Hull:1985jv}. Their result is a correspondence between certain conserved currents associated to symmetries and differential forms $\Phi$ on $\mathcal{M}$ preserved by a connection $\nabla^+$,
\begin{equation}\label{eq:NablaPhi---Intro}
\nabla^+\Phi = 0\,,
\end{equation}
with connection symbols $\Gamma^+=\Gamma+\tfrac{1}{2}\dd B$ twisted by the flux.

In this work we highlight a simple but enlightening generalisation of this result, which we refer to as {\it extended $\mathcal{G}$-structure symmetry}. We assume minimal $(1,0)$ supersymmetry and we include a Fermi sector in the $\sigma$-model, allowing us to incorporate a vector bundle $\mathcal{V}\rightarrow\mathcal{M}$ with gauge field $A$ and curvature $F$, while keeping a general metric and B-field background. We also allow a mass term \cite{AlvarezGaume:1983ab,Hull:1993ct} coupled through a section $S$ of $\mathcal{V}^*$. Our symmetry is described in sections~\ref{ssec:RevHP}--\ref{ssec:GstructureSym}. It holds provided we impose \eqref{eq:NablaPhi---Intro} and further geometric constraints, to be defined and discussed extensively later:
\begin{align} \label{eq:iFPhi---Intro}
&i_F(\Phi)=0\,,
&i_{\dd_A S}(\Phi)=0\, .
\end{align}

We comment on the systems $(\mathcal{V}\rightarrow\mathcal{M}\,;\,G,B,A,S)$ solving these conditions in section~\ref{ssec:Geometry}. These geometries are closely related to supersymmetric backgrounds of heterotic supergravity. Often in this paper we will refer to the natural application of our results to heterotic compactifications. Meanwhile our statements are very general --- we need only minimal supersymmetry --- and valid regardless of the role played by the $\sigma$-model. Hence they are likely to find applications beyond the realm of the heterotic string.

In section~\ref{sec:Anomalies}, we examine whether $\mathcal{G}$-structure symmetries are anomalous. We prove (sect.~\ref{sec:HPSymAnomaly}) that the one-loop quantum effective action corresponding to $(1,0)$ $\sigma$-models \cite{Hull:1986xn} is invariant provided we assign order-$\alpha'$ quantum corrections to the conditions mentioned above. In particular, there must be a connection $\Theta$ on $T\mathcal{M}$ satisfying a curvature condition analogous to \eqref{eq:iFPhi---Intro},
\begin{equation}
i_{R^{\Theta}}(\Phi)=0\,,
\end{equation}
and the torsion in \eqref{eq:NablaPhi---Intro} must be replaced by the gauge-invariant combination
\begin{equation}
\dd B + \frac{\alpha'}{4}\left(\text{CS}_3(A)-\text{CS}_3(\Theta)\right)\,.
\end{equation}
This result is beautifully consistent with the world-sheet Green--Schwarz mechanism in heterotic string theory \cite{Green:1984sg,Hull:1986xn}, reviewed in section~\ref{sec:GreenSchwarz}. This connection appears to have gone unnoticed until now. We discuss how sensitive these results are to counterterm ambiguities.

We also comment on gauge-invariance at order $\alpha'$ in relation with $(1,0)$ superconformal symmetry. Again, with the effective action, we show how to $\alpha'$-correct the supercurrent when flux is turned on. We connect with familiar results on conformal anomalies. Finally, an appendix clarifies classical facts about the superconformal-type chiral symmetries discussed in this paper. The next section sets up our conventions.

\section{Two-dimensional $(1,0)$ non-linear $\sigma$-model} \label{sec:2d NLSM}

\subsection{Conventions} \label{sec:Conventions}

Our $\sigma$-model conventions are as follows \cite{Hull:1985jv,Hull:1986xn,Lambert:1995hs}. We work on a compact world-sheet without boundary in Lorentzian signature and use lightcone coordinates $z^+,z^-$. To avoid cluttering formul\ae, we omit some of the usual Lorentz indices when no confusion is possible. The Grassmann direction is parametrized by $\theta$ and generic  superspace coordinates are denoted by $\zeta^\mu = (z^+,z^-,\theta)$. We write the superspace measure as $d^{2|1}\zeta = \dd z^+ \dd z^- \dd\theta$. The superderivative and supercharge are given by
\begin{equation}
D = \partial_\theta+i\theta\partial_+\, ,\qquad Q = \partial_\theta-i\theta\partial_+\, ,
\end{equation}
where by convention
\begin{equation}
\partial_+z^+=\partial_-z^-=\partial_\theta \theta=1\, .
\end{equation}
They satisfy $-Q^2 = D^2 = i\partial_+$ and both have weights $(h_+,h_-)= (1/2,0)$.

We need two types of superfields,
\be
X^i = x^i +  \theta\psi^i\,,\qquad
\Lambda^\alpha = \lambda^\alpha + \theta f^\alpha\, .
\ee
The Bose superfields $X$ locally define a map $X : \Sigma\,\longrightarrow\, \mathcal{M}$ from superspace $\Sigma$ to a $d$-dimensional target space $\mathcal{M}$, and have weights $(0,0)$. Their leading components are ordinary bosonic fields, while $\psi$ are left-moving Majorana--Weyl fermions. The Fermi superfields $\Lambda$ have weights $(0,1/2)$ and form a section of the bundle $\sqrt{K}_-\otimes {X}^*\mathcal{V}$, where $\mathcal{V}$ is a vector bundle with connection $A$ on the target space $\mathcal{M}$, and $\sqrt{K}$ is the spin bundle over the world-sheet. The Majorana--Weyl fermions $\lambda$ are right-moving and $f$ are auxiliary fields.

The most general renormalisable action preserving $(1,0)$ supersymmetry \cite{Hull:1985jv, Sen:1985qt} that can be written for these fields follows from dimensional analysis. Allowing also for a mass term, we shall consider $\text{S}=\text{S}_{\mathcal{M}}+\text{S}_{\mathcal{V}}+\text{S}_{\mathcal{S}}$, where
\begin{align}
\text{S}_{\mathcal{M}}[X] &= \int_\Sigma\frac{\dd^{2|1}\zeta}{4\pi\alpha'} ~(- i)\, M_{ij}(X)\,  DX^i\partial_-X^j\, , \label{eq:S_M} \\
\text{S}_{\mathcal{V}}[X,\Lambda] &= \int_\Sigma\frac{\dd^{2|1}\zeta}{4\pi\alpha'} ~  \tr ( \Lambda  D_A\Lambda) \, , \label{eq:S_V} \\
\text{S}_{\mathcal{S}}[X,\Lambda] &= \int_\Sigma\frac{\dd^{2|1}\zeta}{4\pi\alpha'} ~  m\, \tr( S(X)\, \Lambda) \, . \label{eq:S_S}
\end{align}
Here $M(X)$ is a $d\times d$ matrix whose symmetric and anti-symmetric parts are the target space metric and Kalb--Ramond field: $M_{ij}=G_{ij}+B_{ij}$. We also use the gauge covariant superspace derivative
\be
D_A\Lambda^\alpha = D\Lambda^\alpha + \hat A^\alpha{}_\beta\Lambda^\beta\, .
\ee
Here and later, we add hats to operators constructed by appending factors of superderivatives of the Bose superfields to expressions with form indices. For example, \be
\hat A^\alpha{}_\beta = A_i{}^\alpha{}_\beta(X) \, D X^i\, .
\ee
The trace over bundle-valued forms is taken with respect to the bundle metric $h_{\alpha\beta}(X)$, so in the expression for the action this means 
\be
\tr(\Lambda D_A\Lambda)  
=  h_{\alpha\beta}(X)\, \Lambda^\alpha D_A\Lambda^\beta\, .
\ee
We choose, without loss of generality, the bundle metric $h_{\alpha\beta}$ to be constant. Finally, $m$ is a \emph{constant} parameter of mass dimension one and $S(X)$ is a section of ${\cal V}^*$. The associated term is a potential for the bosonic fields introduced in \cite{AlvarezGaume:1983ab,Hull:1993ct}. It may be used to cure infrared divergences \cite{AlvarezGaume:1981hn,AlvarezGaume:1983ab} and is related to solitonic effects \cite{Papadopoulos:1994kj} and Landau--Ginzburg theories \cite{Witten:1994tz}.

\subsection{General variations}\label{sec:Vars}

We begin by considering general variations of the action \eqref{eq:S_M}--\eqref{eq:S_S} to prepare the ground for our symmetry,
\begin{equation} \label{eq:deltaS}
\delta \text{S} = \int \frac{\dd^{2|1}\zeta}{4\pi\alpha'} ~ \left(\frac{\delta \text{S}}{\delta X^i}\, \delta X^i + \frac{\delta \text{S}}{\delta \Lambda^\alpha}\, \delta \Lambda^\alpha\right)\,.
\end{equation}
For variations with respect to $X$, we find
\begin{align}
\frac{\delta \text{S}_{\mathcal{M}}}{\delta X^i} &= 2i\, G_{ij}\left( D\partial_- X^j 
+ \Gamma^+{}^{\, j}{}_{kl}\, \partial_-X^k D X^l\right) \, ,\label{eq:EoMX}
\\[3pt]
\frac{\delta \text{S}_{\mathcal{V}}}{\delta X^i} &= \tr\big(\Lambda F_{ij}D X^j\Lambda\big)
+ 2 h_{\alpha\beta}
\big(
 D_A\Lambda^\alpha 
\big) A_i{}^\beta{}_\delta \Lambda^\delta\, ,
\\[3pt]
\frac{\delta \text{S}_{\mathcal{S}}}{\delta X^i} &=  m \, \tr \big((\partial_i S) \Lambda\big)\, ,
\end{align}
where $F$ is the curvature two-form of $A$,
\be
F = \dd A + A\wedge A\, ,
\ee
and we have defined a connection $\nabla^+$  on $T\mathcal{M}$ with symbols $\Gamma^+$ given by
\be
\Gamma^+{}^i{}_{jk} = \Gamma^{\, i}{}_{jk} + \frac{1}{2}\, (\dd B)^i{}_{jk}\, , \label{eq:GammaPlus}
\ee
where $\Gamma$ represents the Levi--Civita connection symbols. In deriving these expressions, we have integrated by parts and discarded boundary terms. We will continue to do so in this paper. The variations with respect to the Fermi superfields are
\be
\frac{\delta \text{S}_{\mathcal{M}}}{\delta \Lambda^\alpha} = 0\, ,
\quad
\frac{\delta \text{S}_{\mathcal{V}}}{\delta \Lambda^\alpha} = 2\, h_{\alpha\beta} \,
 D_A\Lambda^\beta 
\,,
\quad
\frac{\delta \text{S}_{\mathcal{S}}}{\delta \Lambda^\alpha} =  m\, S_\alpha\,.
\ee

It will be easier to demonstrate our symmetry if we write the variations of the $\sigma$-model action these expressions in terms of {\it covariant} perturbations $\delta_A\Lambda$ of $\Lambda$ \cite{Hull:1993ct}, that is
\be
\delta_A \Lambda^\alpha =
\delta\Lambda^\alpha 
+ A_i{}^\alpha{}_\beta \Lambda^\beta\, \delta X^i\,.\label{eq:CovdeltaL}
\ee
In terms of this, a general variation of the action can be written as
\begin{equation} \label{eq:generalCovariantVariationAction}
\delta \text{S} = \int \frac{\dd^{2|1}\zeta}{4\pi\alpha'} \left(\frac{\Delta \text{S}}{\Delta X^i}\, \delta X^i + \frac{\Delta \text{S}}{\Delta \Lambda^\alpha}\, \delta_A \Lambda^\alpha\right)
\, ,
\end{equation}
where we have reorganised the expressions above to define
\begin{align}
\frac{\Delta \text{S}}{\Delta X^i} &= \frac{\delta \text{S}_{\mathcal{M}}}{\delta X^i} + \tr\big(\Lambda F_{ij} D X^j \Lambda\big)
+  m \, \tr \big((\dd_A S)_i\, \Lambda\big)\, , \label{eq:delX}
\\[5pt]
\frac{\Delta \text{S}}{\Delta \Lambda^\alpha} &= 
2\, h_{\alpha\beta}\, D_A\Lambda^\beta 
+ m \, S_\alpha = \frac{\delta \text{S}}{\delta \Lambda^\alpha}
\, .\qquad\qquad\qquad\quad
\label{eq:delLamb}
\end{align}
Here 
\be
\dd_A S = \dd S - SA = (\partial_i S - S A_i)\, \dd x^i~
\ee
is the appropriate covariant exterior derivative for the  section $S$ of ${\cal V}^*$.

\section{Extended $\mathcal{G}$-structure symmetry} \label{sec:Extended G-structure symmetry}

Suppose the target space manifold $\mathcal{M}$,  with ${\rm dim}\, {\cal M} = d$, admits a globally-defined nowhere-vanishing $p$-form $\Phi$. The existence of such a form amounts to a reduction of the structure group $GL(d)$ of the frame bundle of $\cal M$ to a subgroup $\cal G$.  
Howe and Papadopoulos \cite{Howe:1991ic,Howe:1991vs,Howe:1991im} showed that if $\Phi$ satisfies a certain constraint, then the $\sigma$-model with action $\text{S}_{\cal M}$ (that is, in the case where $\Lambda = 0$) has an extra symmetry.
In this section we generalise this symmetry to include the bundle and the mass terms by $\text{S}_\mathcal{V}$ and $\text{S}_\mathcal{S}$. That is, we extend $\mathcal{G}$-structure symmetries to the full non-linear $\sigma$-model.

\subsection{Review of the Howe--Papadopoulos symmetry}\label{ssec:RevHP}

Let $\epsilon(\zeta)$ be a general function over superspace with left-moving weight $h_+=(1-p)/2$. It has even/odd Grassmann parity depending on whether $p$ is odd/even.
Consider the transformation
\begin{align} \label{eq:Howe-Pap}
&\delta^{\Phi} X^i = \,
\frac{\epsilon(\zeta)}{(p-1)!}\, 
\,\Phi^i{}_{i_2\ldots i_p}(X) D X^{i_2\ldots i_p}
= \epsilon(\zeta)\, \hat\Phi^i\,,
\, 
\end{align}
where $DX^{i_2\ldots i_p}$ is a shorthand for $DX^{i_2}\ldots DX^{i_p}$.

The variation of the $\sigma$-model action $\text{S}_{\cal M}$ induced by \eqref{eq:Howe-Pap} follows from the analysis of section \ref{sec:Vars}. Only the first terms in \eqref{eq:generalCovariantVariationAction} and in \eqref{eq:delX} participate and after integrating by parts we find
\be
\delta \text{S}_{\cal M} = \int \frac{\dd^2z}{4\pi\alpha'}\, \epsilon(\zeta) \dd\theta
\, (-2i) \nabla^+_i\hat\Phi\,\partial_- X^i   
+ \  \int_\Sigma \frac{\dd^2z}{4\pi\alpha'}~ \partial_-\epsilon(\zeta)\,  \dd\theta\,  \hat \Phi\, , 
\label{eq:HPdeltaSM}
\ee
where we have defined
\be
\hat \Phi =  \frac{1}{p}\, D X^i\,  \hat\Phi_i = \frac{1}{p!}\,  \Phi_{i_1\cdots i_p}\,  D X^{i_1\cdots i_p}\,
\ee
and\footnote{We take covariant derivatives to act as $\nabla_i\Phi_j=\partial_i\Phi_j-\Gamma^k{}_{ij}\Phi_k$ on one forms.}
\be
\nabla^+_i\hat\Phi = 
\frac{1}{p!}\, \nabla^+_i\Phi_{j_1\ldots j_p}\, DX^{j_1\ldots j_p} \,.
\ee
It is easy to see that the first term of \eqref{eq:HPdeltaSM} gives the main result of \cite{Howe:1991ic}. If $\Phi$ is parallel under the connection $\nabla^+$ with torsion\footnote{The torsion $T$ of a covariant derivative $\nabla$ which has connection symbols $\Gamma^i{}_{jk}$ is defined as $T^i{}_{jk} = 2 \Gamma^i{}_{[jk]}$.} $T=\dd B$, i.e.
\begin{equation}
\nabla^{+}_i\Phi_{j_1j_2\ldots{j_p}}=0\qquad \text{(} T = \dd B \text{)}\,,
\end{equation}
then the first term vanishes identically. Moreover, if $\epsilon=\epsilon(z^+,\theta)$ is purely left-moving, the last term in equation \eqref{eq:HPdeltaSM} also vanishes and we conclude that \eqref{eq:Howe-Pap} is an infinitesimal chiral symmetry of ${\text S}_\mathcal{M}$. Note that this symmetry is non-linear for $p\geq 3$. The last term in equation \eqref{eq:HPdeltaSM} corresponds to the current \cite{Howe:1991ic}
\begin{equation} \label{eq:HPcurrent}
J^- = 2i(-1)^{p}\hat {\Phi}\, , 
\end{equation}
which is simply the operator naturally associated to the differential form $\Phi$. Its conservation equation is the statement that, up to the equation of motion for $X$, it is left-moving:
\begin{equation}
\partial_-\hat{\Phi} \approx 0\,.
\end{equation}
General facts and notations about chiral symmetries can be found in the appendix, in particular our normalisation leading to the overall factor in \eqref{eq:HPcurrent}.

Chiral currents such as \eqref{eq:HPcurrent} are fundamental to the quantum description of conformal field theories in terms of $\mathcal{W}$-algebras. The $p$-form current here and the classical stress-tensor derived in the appendix (c.f.\ equation \eqref{eq:T+}) are classical limits of corresponding generators in appropriate $\mathcal{W}$-algebras. (We give examples in section 3.2.) An importance difference between them to bear in mind is that a choice of normal ordering must be made when quantum operators are build from classical composite fields. In interacting theories, such as our non-linear $\sigma$-models, there is, to our knowledge, no canonical resolution to this ordering ambiguity.

\subsection{The extended $\cal G$-structure symmetry} \label{ssec:GstructureSym}
We now proceed to generalise the Howe--Papadopoulos symmetry to the full model, including the gauge and mass sectors. As a
first step, we set $\delta_A\Lambda = 0$ and focus on the variation of $\text{S}$ induced only by the Howe--Papadopoulos transformation of $X$ given by \eqref{eq:Howe-Pap}. The ansatz $\delta^\Phi_A\Lambda = 0$ will be relaxed below.\footnote{Recall that the covariant variation of $\Lambda$ is given in 
equation \eqref{eq:CovdeltaL}. Nevertheless $\delta^\Phi_A\Lambda^\alpha=0$ means that $\Lambda$ does transform according to  $\delta^\Phi\Lambda^\alpha = -A_i{}^\alpha{}_\beta\Lambda^\beta \delta^\Phi X^i$.}
All the terms in \eqref{eq:delX} now participate and we find, for the full $\sigma$-model variation,
\begin{equation}
\begin{split}
\delta \text{S} &= \int \frac{\dd^2z}{4\pi\alpha'}\, \epsilon(\zeta) \dd\theta
\left( 
(-2i) \nabla^+_i\hat\Phi\,\partial_- X^i  
+   \tr\Big(
\Lambda F_{ij}\, D X^j\hat \Phi^i \Lambda
+ (-1)^{p-1} m(\dd_A S)_i\,\hat \Phi^i \Lambda\Big)
\right)
\\[3pt]
&\qquad\qquad
+ \  \int_\Sigma \frac{\dd^2z}{4\pi\alpha'}~ \partial_-\epsilon(\zeta)\,  \dd\theta\,  (-2i)\hat \Phi\,.
\end{split} \label{eq:HPdelSX}\raisetag{20pt}
\end{equation}

The vanishing of the first term in \eqref{eq:HPdelSX} gives back of course the results reviewed in section \ref{ssec:RevHP}, but there are now two extra terms. As it is manifest in \eqref{eq:HPdelSX}, we have a symmetry if the following geometric conditions are satisfied
\begin{align}
&\nabla^{+}_i\Phi_{j_1j_2\ldots{j_p}}=0\qquad \text{(} T = \dd B \text{)}\,,\label{eq:HPcondition} \\ 
&F_{i[j_1}\Phi^i{}_{j_2\ldots j_p]}=0\, , \label{eq:Fisetcondition}\\
& (\dd_A S)_i\, \Phi^i{}_{j_2\ldots j_{p}} = 0\, . \label{eq:Mcondition}
\end{align}

As we will see, this already constitutes an interesting extension of the Howe--Papadopoulos symmetry, but we can generalise it one step further. We keep \eqref{eq:Howe-Pap}, but we now also assign a covariant variation to the Fermi superfields:
\begin{align}
&\delta^{\Phi} X^i = \frac{\epsilon(\zeta)}{(p-1)!}\,\Phi^i{}_{i_2\ldots i_p}(X) D X^{i_2\ldots i_p}
= \epsilon(\zeta)\, \hat\Phi^i\,, \label{eq:NonZerodelX} \\
&\delta_A^\Phi\Lambda^\alpha  =  \epsilon(\zeta)\hat\Upsilon^\alpha{}_\beta\, (2 D_A\Lambda^\beta + m\, S^\beta)  =  \epsilon(\zeta)\hat\Upsilon^\alpha{}_\beta \frac{\Delta{\text S}}{\Delta \Lambda_\beta}\,.\label{eq:NonZerodelL}
\end{align}
Here, a priori, the superfield
\begin{align}
\hat\Upsilon^\alpha{}_\beta\,
=\frac{1}{(p-2)!}\, 
\,\Upsilon^\alpha{}_\beta{}_{i_1\ldots i_{p-2}}(X) D X^{i_1\ldots i_{p-2}} \label{eq:Ups}
\end{align}
corresponds to an arbitrary $\text{End}(\mathcal{V})$-valued differential $(p-2)$-form. It is easy to see from \eqref{eq:generalCovariantVariationAction} that the variation $\delta \text{S}$ corresponding to \eqref{eq:NonZerodelX}--\eqref{eq:NonZerodelL} is composed of \eqref{eq:HPdelSX} as well as the extra term
\begin{equation}
\int \frac{\dd^{2|1}\zeta}{4\pi\alpha'} ~ \frac{\Delta \text{S}}{\Delta \Lambda^\alpha}\, \delta_A^\Phi \Lambda^\alpha = \int \frac{\dd^{2|1}\zeta}{4\pi\alpha'} ~ \epsilon(\zeta)\hat{\Upsilon}_{\alpha\beta}\frac{\Delta \text{S}}{\Delta \Lambda_\alpha}\frac{\Delta \text{S}}{\Delta \Lambda_\beta}
\,.\label{eq:HPdelSL}
\end{equation}
The vanishing of this term is achieved if the endomorphism-valued form satisfies $\Upsilon_{(\alpha\beta)}=0$, in other words, whenever $\Upsilon\in\Omega^{p-2}(\mathcal{M},\bigwedge^2\mathcal{V})$. Summarising,

\vspace{5pt}

\noindent\textit{
\eqref{eq:NonZerodelX}--\eqref{eq:NonZerodelL} is a symmetry of the full $\sigma$-model \eqref{eq:S_M}--\eqref{eq:S_S} if the geometric conditions \eqref{eq:HPcondition}--\eqref{eq:Mcondition} and $\Upsilon_{(\alpha\beta)}=0$ are satisfied.
}

\vspace{5pt}

This transformation was in fact considered in \cite{Howe:1988cj,Hull:1993ct} in the case ${\cal G} = U(d/2)$ and $p= 2$.  In these references, the Howe--Papadopoulos symmetry is constructed such that it is a new supersymmetry transformation hence enhancing the superconformal symmetry to $(2,0)$. In this case, the form $\Upsilon$ is a section of ${\rm End}({\cal V})$ and it corresponds to a complex structure on $\cal V$.

The constraints needed for extended $\mathcal{G}$-structure symmetries, \eqref{eq:Fisetcondition} and \eqref{eq:Mcondition},  can be written nicely in terms of insertion operators. An insertion operator is a linear map which satisfies the Leibniz rule, that is, it is a derivation, which is defined as follows. Consider the space of forms, perhaps with values in a vector bundle $\cal E$ over $\cal M$ which we denote as $\Omega^\bullet({\cal M}, {\cal E})$. Let $P$ be a $p$-form with values in the tangent bundle of $\cal M$, that is $P\in \Omega^p({\cal M}, T{\cal M})$.  The insertion operator $i_P$ is a derivation on  $\Omega^\bullet({\cal M}, {\cal E})$ of degree $p-1$  defined by
\be
\begin{split}
i_P~:~ \Omega^k({\cal M}, {\cal E}) &~~\longrightarrow~~ \Omega^{k+p-1}({\cal M}, {\cal E})
\\
\alpha ~~\quad&~~~\mapsto ~~~ i_P(\alpha) = P^i\wedge\alpha_i\, , 
\end{split}\label{eq:insert}
\ee
where $\alpha$ is any $k$-form and 
\be
\alpha_i = \frac{1}{(k-1)!}\, \alpha_{ij_1\cdots j_{k-1}}\, 
\dd x^{j_1\cdots j_{k-1}}\,.
\ee

The constraint equation \eqref{eq:Fisetcondition}, which restricts the connection $A$ on the bundle $\cal{V}$,   can be written as
\be
i_F(\Phi) = 0 \,,\label{eq:InstGeneral}
\ee
where in this equation $F$ is interpreted as a one form with values in ${T{\cal M}\otimes\rm End}({\cal V})$,
\be
F^i = G^{ij}\, F_{jk}\, \dd x^k\,.
\ee 
In this paper we say that a connection $A$ which satisfies this condition is a {\it $\sigma$-model quasi-instanton}.   As we illustrate in section \ref{ssec:Geometry}, in some {\it definite} examples this condition does agree on the nose with the usual notion of a gauge bundle instanton in heterotic supergravity. 
More generally however, we do not have this equivalence.

Similarly, using insertion operators, the constraint \eqref{eq:Mcondition} can  be written as
\be
i_{\dd_A S}(\Phi) = 0 \,.\label{eq:iMcondition}
\ee

In summary, the conditions for our extended $\mathcal{G}$-structure symmetry to hold are written as
\be
\nabla^+\, \Phi = 0\, , \quad i_F(\Phi)= 0\,,
\quad
i_{\dd_A S}(\Phi) = 0\, , \quad \Upsilon_{(\alpha\beta)} = 0\,.\label{eq:AllConditions}
\ee
In section \ref{sec:Anomalies} we consider the potential anomalies of this symmetry and show that the one-loop effective action is invariant as long as we assign appropriate $\alpha'$-corrections to these conditions.

We return below to a description of these geometric constraints. Before doing so, it is worth noting the following remarkable fact:

\vspace{5pt}

\noindent\textit{The conserved current for the extended $\mathcal{G}$-structure symmetry is the same as for the corresponding Howe--Papadopoulos symmetry: the bundle sector and the mass terms do not affect the current.}

\vspace{5pt}

\noindent This follows from \eqref{eq:HPdelSX} and \eqref{eq:HPdelSL}. In the classical limit, this fact explains why the bundle sector does not feature prominently in abstract conformal field theoretic descriptions of heterotic compactifications. An analysis based only on currents can hardly distinguish between the models with and without bundles.

To illustrate this fact, we mention \cite{Melnikov:2017yvz}, where the authors identify the internal superconformal algebras preserving various amounts of supersymmetry in Minkowski space-times of low dimensions $10-d$ after compactifying critical heterotic string theory. Focusing on minimal space-time supersymmetry, they find the so-called $\mathcal{SW}(3/2,2)$ algebra at $c=12$ in the case $d=8$ and an algebra of type $\mathcal{SW}(3/2,3/2,2)$ with $c=21/2$ for $d=7$. These two algebras were originally introduced in the context of type II string compactifications \cite{Shatashvili:1994zw} on $Spin(7)$ and $G_2$ holonomy manifolds respectively. Obviously no vector bundles arise in type II, but the associated $W$-algebras nevertheless play a role in heterotic strings.

Another heterotic application of the $\mathcal{SW}(3/2,3/2,2)$ algebra for $G_2$ features in \cite{Fiset:2017auc}. The algebra was used to define a world-sheet BRST operator whose cohomology contains infinitesimal marginal deformations of the conformal field theory. Again, the heterotic vector bundle was encompassed almost automatically in the framework.

\subsection{Geometrical constraints on $(\cal M, \cal V)$}\label{ssec:Geometry}

As discussed earlier,  the existence of a well-defined nowhere-vanishing $p$-form $\Phi$ on the target space $\mathcal{M}$ of dimension $d$ amounts to a reduction of the structure group  to ${\cal G}\subset GL(d)$. Interesting examples are $SO(d)$, $U(d/2)$, $SU(d/2)$, $Sp(d/4)$, $Sp(1)\cdot Sp(d/4)$, $G_2$, and $Spin(7)$. In some cases (as for example $Spin(7)$, $G_2$ and $SU(d/2)$) the target manifold admits at least one well-defined nowhere-vanishing spinor which is the basic topological condition on the target manifold necessary to obtain space-time supersymmetric effective field theories. 

We now turn to a geometrical explanation of the conditions \eqref{eq:AllConditions}.   We illustrate these with examples of Riemannian target spaces so that ${\cal G}\subset SO(d)$, and which are related to heterotic supergravity compactifications preserving at least one space-time supersymmetry.

Consider, for example, an eight dimensional target space with structure group 
$\mathcal{G} = Spin(7)$, or a seven dimensional manifold with ${\cal G} = G_2$.
In both of these cases, the form $\Phi$ is of degree $p=4$, and such target spaces admit one well-defined nowhere-vanishing spinor.  Compactifying heterotic supergravity on a manifold with a $G_2$-structure gives rise to three dimensional Yang--Mills  $N=1$ supergravity \cite{Gunaydin:1995ku,Gauntlett:2001ur,Friedrich:2001nh,Friedrich:2001yp,Gauntlett:2003cy,Ivanov:2003nd}, while compactification on a manifold with a $Spin(7)$ structure gives a $(1,0)$ supersymmetric two dimensional field theory \cite{Ivanov:2003nd,Ivanov:2001ma}. 

Other interesting examples are $\mathcal{G}=U(d/2)$ and $\mathcal{G}=SU(d/2)$.  The case where $\mathcal{G}=U(d/2)$ corresponds to even dimensional almost Hermitian target spaces where $\Phi=\omega$ is the Hermitian two form. Heterotic string compactifications on almost Hermitian manifolds are not supersymmetric (the group $U(d/2)\subset SO(d)$ does not leave any invariant spinors) unless the structure group is reduced further to $G=SU(d/2)$.  In this case, there is another nowhere-vanishing form $\Phi= \Omega$ with $p =n=d/2$ and the corresponding target spaces are almost Hermitian with vanishing first Chern class. Compactifying heterotic supergravity on a manifold with such an $SU(d/2)$-structure is not yet sufficient to obtain a supersymmetric space-time supergravity.  For example, it was shown in \cite{Hull:1986kz,Strominger:1986uh} that when $d=6$,  one needs to demand further that the almost complex structure is integrable to obtain space-time Yang--Mills  $N=1$ supergravity.  Furthermore, as mentioned in section \ref{ssec:GstructureSym}, the Howe-Papadopoulos symmetry in \cite{Howe:1988cj,Hull:1993ct} corresponds precisely to an enhancement of the superconformal symmetry to $(2,0)$.  This is of course beautifully consistent with the work of \cite{Banks:1987cy} in which it is shown that the world-sheet quantum field theory corresponding to a four dimensional supersymmetric space-time theory obtained from superstring compactifications must be  $N=2$ superconformal invariant.

Manifolds with a $\mathcal{G}$-structure admit connections $\nabla$ which are metric and are compatible with the  $\mathcal{G}$-structure, that is $\nabla \Phi = 0$. These connections have an {\it intrinsic} torsion $T(\Phi)$ which is uniquely determined by the $\mathcal{G}$-structure $\Phi$. Equation \eqref{eq:HPcondition} says that the form $\Phi$ needs to be covariantly constant with respect to a connection with totally antisymmetric torsion $T(\Phi) = \dd B.$
Note that this relation ties the target space geometry with the physical flux.
Not all manifolds with a given $\mathcal{G}$-structure admit such a connection with a totally antisymmetric torsion except in the case of ${\cal G} = Spin(7)$ \cite{Ivanov:2001ma}.
For instance, when $\mathcal{G} = G_2$, taking $\Phi$ to be the co-associative four form, the necessary and sufficient condition for the existence of a $G_2$-compatible connection with totally antisymmetric torsion is that the five form $\dd \Phi$ is in the ${\bf 7}$ dimensional representation\footnote{A five form on a manifold with a $G_2$ structure decomposes into the $G_2$ irreducible representations {\bf 7}+{\bf 14}.} of $G_2$. In fact, in this case there is a unique $G_2$-compatible connection with totally antisymmetric torsion \cite{Bryant:2005mz}. In even dimensions, with $\mathcal{G}=U(d/2)$, there exists a unique metric connection compatible with the $U(d/2)$ structure with totally antisymmetric torsion which is called the {\it Bismut} connection \cite{MR1006380, Friedrich:2001nh}.\footnote{Note that the complex structure does not need to be integrable for this statement to be true.}

We now turn to the $\sigma$-model quasi instanton connection $A$ on the bundle $\cal V$
\be
i_F(\Phi) = 0 \,.\label{eq:InstGeneral2}
\ee
For the examples pertaining to the heterotic compactifications with ${\cal G} = Spin(7), G_2$ or $SU(n)$, we want to see to what extent this corresponds to the instanton condition obtained from the BPS equations in heterotic supergravity, in particular, to the vanishing of the supersymmetric variations of the gaugino.

Suppose the target manifold admits a $Spin(7)$ or  a $G_2$ structure.
It is a well known fact about the geometry of these manifolds that \eqref{eq:InstGeneral2} is equivalent to $F\in\Omega^2_{21}({\cal M}, {\rm End}({\cal V}))$
in the case of $Spin(7)$, and to $F\in\Omega^2_{14}({\cal M}, {\rm End}({\cal V}))$ for $G_2$ \cite{MR2164593}. In both cases, one can in fact write this condition using an appropriate projection operator on $F$ into the appropriate irreducible representation of $\cal G$.
\be
(2\delta^{kl}_{ij} +  \Phi^{kl}{}_{ij}) F_{kl} = 0\, ,
\ee
or equivalently,
\be
F\lrcorner \Phi = - F\, . \label{eq:Inst78}
\ee


In the case of $U(n)$ structures (where the dimension of $\cal M$ is $d = 2n$)\footnote{In this case, there always exists an almost complex structure $J$  with $J^2 = - {1}$, and conversely, the existence of an almost complex structure on a Riemannian manifold reduces the structure group to $U(n)$.}
the condition \eqref{eq:InstGeneral2} constrains the bundle $\cal V$ to be holomorphic, that is, we have
\be
i_F(\omega) = 0\, \quad\iff\quad F^{(0,2)} = 0\,.\label{eq:InstU(n)}
\ee
To see this, note that 
\be
i_F(\omega) = F^i\wedge\omega_i = - F_{ki}J^k{}_j \dd x^{ij}\,, 
\qquad J^i{}_j = G^{ik}\omega_{jk}\, ,
\ee
where $J$ is the almost complex structure. 
Therefore, 
\be
i_F(\omega)= 0~\iff~F_{ij} =  J^k{}_i J^l{}_j\, F_{kl}\, ,
\ee
and the result \eqref{eq:InstU(n)} follows.

If moreover, the structure group reduces to $SU(n)$, then there is a further constraint on $\cal V$ due to the existence of a second $n$-form $\Omega$, which is
\be
 i_F(\Omega) = 0\,\quad\iff\quad \omega\lrcorner F = 0\, . \label{eq:InstSU(n)}
\ee
that is,  $F$ must be a \textit{primitive} two form.
To see this equivalence, note first that, when $F^{(0,2)} = 0$, the three form  $i_F(\Omega)$ must be type $(n,0)$. Then 
\be
i_F(\Omega)= 0~\iff~\overline\Omega\lrcorner\, i_F(\Omega) = 0
\, . 
\ee
Noting that the $(n,0)$  form $\Omega$
satisfies
\be
\overline\Omega^{ik_1\cdots k_{n-1}}\,\Omega_{jk_1\cdots k_{n-1}}
\propto \delta^i{}_j - i\, J^i{}_j\, , 
\ee
we obtain 
\be
i_F(\Omega)= 0~\iff~\overline\Omega\lrcorner\, i_F(\Omega) = 0
~\iff~ \omega\lrcorner F = 0\, . 
\ee

Together, the conditions \eqref{eq:InstU(n)} and \eqref{eq:InstSU(n)}, are equivalent to  $F$ being a primitive $(1,1)$ form or, equivalently, $F\in \Omega_{\bf adj}^2({\cal M}, {\rm End}({\cal V}))$.

In the $SU(3)$ case, one can also show that the last condition is equivalent to 
\be
F\lrcorner \, \rho = -F\,,\qquad \rho= \frac{1}{2}\, \omega\wedge\omega\, .
\ee
Note the similarity with equation \eqref{eq:Inst78}.

In summary, for the examples $\mathcal{G}= Spin(7), G_2$ and $SU(3)$, we have that
\be
F\in \Omega^2_{\bf adj} ( {\cal M}, {\rm End}({\cal V}))\, ,
\ee
where $\bf adj$ is the adjoint representation of $\mathcal{G}$, or equivalently
\be
F\lrcorner \Phi = - F\, ,
\ee
where $\Phi$ is the Cayley four form for $\mathcal{G}= Spin(7)$ structure, the co-associative four form for $\mathcal{G}= G_2$, and, for $\mathcal{G}= SU(3)$, we have $\Phi = \rho$.  We say that the bundle connection $A$ satisfying this condition is an {\it instanton}, meaning that the curvature $F$ of such an instanton connection $A$ on the bundle satisfies the Yang--Mills equation\footnote{See for example the paper by Harland and N\"olle \cite{Harland:2011zs} which contains a very good discussion about instantons as solutions of the Yang--Mills equation.} 
\be
\dd_A^\dagger F = - F\lrcorner \dd^\dagger \Phi\,.\label{eq:GenInst}
\ee

The final constraint \eqref{eq:iMcondition} can be written as
\be
\dd_A S = 0\, ,\label{eq:flatS}
\ee
for the examples at hand, because $\dd_A S$ is a one form. This means that the section $S$ must be a flat section. In this paper however we will not be concerned with this condition any further.

We close here with a comment about the minimally supersymmetric  heterotic compactifications we have been discussing in this section. In order for these supersymmetric solutions to satisfy the supergravity equations of motion to first order in $\alpha'$, it is also necessary that there is a connection $\Theta$ on the tangent bundle $T{\cal M}$ which is an instanton \cite{Ivanov:2009rh}. On the other hand, in the $(1,0)$ $\sigma$-model the connection $\Theta$ appears in order to cancel the gravitational anomalies. We will see in the following section that to first order in $\alpha'$,  the 
one-loop effective action is invariant under the $\cal G$-structure symmetries, provided that $\Theta$ is a $\sigma$-model quasi-instanton together with the usual corrections to the torsion involving the Chern--Simons forms for $A$ and $\Theta$.

\section{Anomalies} \label{sec:Anomalies}

In this part, we propose an analysis of anomalies of $\mathcal{G}$-structure and superconformal symmetries from the angle of effective actions.

In the absence of anomalies, a standard argument shows formally that the $\sigma$-model effective action \cite{Howe:1986vm,Hull:1986xn,Hull:1986hn}, denoted $\Upgamma$, obeys the Slavnov--Taylor identity
\begin{equation} \label{eq:SlavnovTaylor}
\int \frac{d^{2|1}\zeta}{4\pi\alpha'}\left(\frac{\delta \Upgamma[X,\Lambda]}{\delta X^i}\langle{ \overline{\delta X^i} }\rangle + \frac{\delta \Upgamma[X,\Lambda]}{\delta \Lambda^\alpha}\langle{ \overline{\delta \Lambda^\alpha} }\rangle\right)=0\,,
\end{equation}
where the expectation values are taken with background sources for the dynamical fields, and where the bars refer to specific symmetry variations as explained near \eqref{eq:BarSymmetry1}--\eqref{eq:BarSymmetry2}. In the presence of chiral fermions, the functional measure in the path integral generically transforms anomalously, leading to a non-vanishing right hand side in \eqref{eq:SlavnovTaylor}. For linear symmetries, this can be probed by a first order variation of the effective action since $\langle{ \overline{\delta X^i} }\rangle = \overline{\delta X^i}$ and similarly for the Fermi superfield.

General $\mathcal{G}$-structure symmetries are however non-linear which complicates their analysis. Very few research efforts appear to have gone into this problem; in fact, we are only aware of \cite{Howe:2006si}. In this work, anomalies of $\mathcal{G}$-structure symmetries of $(1,1)$ models without torsion ($\dd B = 0$) were examined within a BV--BRST framework, and multiple related difficulties were highlighted.

In this paper, we take a simplified approach which nevertheless yields results consistent with supergravity.
We assume an expansion in powers of $\alpha'$ of the form
\begin{equation}
\langle{ \overline{\delta X^i} }\rangle = \overline{\delta X^i}+\alpha'\overline{\delta X^i_{(1)}}+O(\alpha'{}^2)
\end{equation}
and address anomalies perturbatively. We allow for the possibility of $\alpha'$-corrections to $\mathcal{G}$-structure symmetries.
We also use an effective action computed order by order using the background field method \cite{DeWitt:1967uc,Honerkamp:1971sh,AlvarezGaume:1981hn,Braaten:1985is,Henty:1987wc,Gates:1986ez}. This implies the usual limitations: target space curvature and fluxes must be small and slowly varying in string units.\footnote{Allowing non-trivial fluxes makes this assumption questionable and it is important to verify self-consistency.}

Our method is analogous to the treatment of sigma-model anomalies \cite{Moore:1984dc,AlvarezGaume:1985yb,Bagger:1985pw} for target space gauge and Lorentz transformations \cite{Hull:1985jv,Sen:1985tq,Sen:1986nm}, especially as covered in \cite{Hull:1986xn}. We start by reviewing this discussion in order to prepare the ground for our treatment of $\mathcal{G}$-structure anomalies in section~\ref{sec:HPSymAnomaly}. The corresponding analysis of conformal anomalies is then presented for comparison in section~\ref{sec:Superconformal anomalies}.

\subsection{Effective action and Green--Schwarz mechanism} \label{sec:GreenSchwarz}

In 1986, there were some questions as to whether the world-sheet implementation of the target space Green--Schwarz cancellation mechanism \cite{Green:1984sg,Hull:1985jv} was consistent with $(1,0)$ supersymmetry \cite{Sen:1985tq, Sen:1986nm}. Hull and Townsend addressed the issue in \cite{Hull:1986xn} by calculating a world-sheet one-loop effective action directly in superspace and used it to cancel the anomaly supersymmetrically. They found that the non-local
and gauge non-invariant contributions can be packaged as
\begin{equation} \label{eq:San}
\text{S}_{\text{an}}^{(A)}[X]=\int \frac{d^{2|1}\zeta}{4\pi\alpha'}~\frac{\alpha'}{4}\text{tr}\left( \partial_- \hat{A} \,\frac{1}{\partial_+}D\hat{A}\right) ,
\end{equation}
where $\hat{A}=A_i(X)DX^i$ and where the trace is over the gauge indices of $A$. There is also a similar term due to gravitational anomalies,
\begin{equation} \label{eq:SanTheta}
\text{S}_{\text{an}}^{(\Theta)}[X]=\int \frac{d^{2|1}\zeta}{4\pi\alpha'} ~(-1)\, \frac{\alpha'}{4}\text{tr}\left(\partial_- \hat{\Theta}\,  \frac{1}{\partial_+}D\hat{\Theta}\right),
\end{equation}
where $\Theta$ is the spin connection on $T\mathcal{M}$ associated to a covariant derivative $\nabla^-$ with connection symbols $\Gamma^-$  defined by\footnote{We remark that $\nabla^-$ is not compatible with the $\cal G$-structure ($\nabla^-\Phi \ne 0$).  It is however metric because the torsion is totally antisymmetric.} 
\be
\Gamma^-{}^i{}_{jk} = \Gamma^i{}_{jk} - \frac{1}{2}\, (\dd B)^i{}_{jk}\, ,\label{eq:GammaMinus}
\ee
and where we must now trace over the corresponding Roman spin indices. 
The analysis of $\text{S}_{\text{an}}^{(\Theta)}$ is entirely parallel to that of $\text{S}_{\text{an}}^{(A)}$ so we will mostly omit it for simplicity.

The formal inverse in \eqref{eq:San} is defined using the Green's function of $\partial_+$. We refer to \cite{Polchinski:1998rq} for more details. For our purposes it suffices to know that an analogue of integration by parts holds for such operators, and that it can be commuted through ordinary differential operators.

We stress that $\text{S}_{\text{an}}^{(A)}$ correctly describes gauge non-invariance only up to quadratic order in the gauge connection. This is important for example when describing the Green--Schwarz mechanism, which we now review. Under a general variation of $\hat{A}$, $\text{S}_{\text{an}}^{(A)}$ transforms as
\begin{equation} \label{eq:deltaSan}
\delta \text{S}_{\text{an}}^{(A)}=  \int \frac{d^{2|1}\zeta}{4\pi\alpha'} ~ (-1)\, \frac{\alpha'}{2} \text{tr} \left[\left(\partial_-\frac{1}{\partial_+}D\hat{A}\right)\delta \hat{A}\right].
\end{equation}
In order to check gauge anomalies, we substitute in~\eqref{eq:deltaSan} the target space gauge variation
\begin{equation}
\updelta \hat{A} = \big(\partial_i\Xi+[A_i,\Xi]\big)DX^i = D\Xi+[\hat{A},\Xi]
\end{equation}
where we use $\updelta$ to distinguish gauge from generic variations and where $\Xi(X)$ is the gauge parameter. We obtain
\begin{equation}
\updelta \text{S}_{\text{an}}^{(A)}=- i \int \frac{d^{2|1}\zeta}{4\pi\alpha'} ~ \frac{\alpha'}{2} \text{tr}\left( A_i(\partial_j\Xi) \,DX^i\partial_- X^j\right),
\end{equation}
which is finite and local. Clearly, this has the same form as the classical action \eqref{eq:S_M} and can thus be cancelled by assigning a compensating gauge variation to $M_{ij}$,
\begin{equation}
\updelta M_{ij}(X) = - \frac{\alpha'}{2}\text{tr}\left(A_i\,\partial_j\Xi \right)\,.
\end{equation}

This process cancels the anomaly but introduces a gauge-\textit{variant} metric. One generally prefers to work with invariant objects. This can be achieved as follows. The effective action is only well-defined up to finite local counterterms. Consider then adding the metric counterterm
\begin{equation} \label{eq:Sadd}
\text{S}_{\text{add}}^{(A)}[X]=i\int \frac{d^{2|1}\zeta}{4\pi\alpha'}~\frac{\alpha'}{4}\text{tr}(\hat{A}A_j\partial_-X^j)\,.
\end{equation}

Under a gauge transformation, this varies by
\begin{equation}
\updelta \text{S}_{\text{add}}^{(A)}=i\int \frac{d^{2|1}\zeta}{4\pi\alpha'}~\frac{\alpha'}{2} \text{tr}\left(A_{(i}\partial_{j)}\Xi )\right) DX^i\partial_- X^j\,.
\end{equation}
This is enough to cancel the symmetric part of the anomaly,
\begin{equation}
\updelta (\text{S}_{\text{an}}^{(A)}+\text{S}_{\text{add}}^{(A)})= - i\int \frac{d^{2|1}\zeta}{4\pi\alpha'}~\, \frac{\alpha'}{2}\text{tr}\left(\Xi\partial_{[i}A_{j]}\right)DX^i\partial_- X^j\,,
\end{equation} 
so that we may simply assign an anomalous gauge transformation to the B-field,
\begin{equation}
\updelta B_{ij}= - \frac{\alpha'}{2}\,\tr\left(\Xi\,\partial_{[i}A_{j]}\right)\,,
\end{equation}
and leave the metric gauge-invariant. Repeating this argument for \eqref{eq:SanTheta} we obtain the usual Green--Schwarz mechanism and heterotic Bianchi identity. The associated gauge-invariant field strength involves Chern--Simons three forms for the gauge and tangent bundle connections:
\begin{equation} \label{eq:H}
H=\text{d}B+ \frac{\alpha'}{4}\left(\text{CS}_3(A)-\text{CS}_3(\Theta)\right).
\end{equation}
We briefly note that there is an ambiguity in this cancellation scheme, whereby any connection $\Theta$ can be used in the Bianchi identity. This ambiguity is usually lifted by demanding conformal symmetry at the quantum level. For more details, we refer to \cite{Hull:1985dx}. As we discuss in the next part, it can also be fixed by demanding preservation of our extended $\mathcal{G}$-structure symmetries at first order in $\alpha'$.

\subsection{$\mathcal{G}$-structure symmetries and $\alpha'$-corrections} \label{sec:HPSymAnomaly}

We now investigate the effects of a $\mathcal{G}$-structure transformation on $\text{S}_{\text{an}}^{(A)}+\text{S}_{\text{add}}^{(A)}$. The gauge field varies here purely through the chain rule,
\begin{equation}
\delta^\Phi \hat{A} = \partial_iA_j(\delta^\Phi X^i)DX^j + A_iD(\delta^\Phi X^i).
\end{equation}
Substituting this in \eqref{eq:deltaSan}, we find the nonlocal result
\begin{equation} \label{eq:deltaSanBasic}
\delta^\Phi \text{S}_{\text{an}}^{(A)}= \int \frac{d^{2|1}\zeta}{4\pi\alpha'} ~ \frac{\alpha'}{2} \text{tr} \left[-2\left(\frac{\partial_-D}{\partial_+}\hat{A}\right)\partial_{[i} A_{j]}DX^j+i(\partial_- \hat{A}) A_i\right]\delta^{\Phi}X^i.
\end{equation} 
Next we vary the local counterterm. We obtain
\begin{equation}
\begin{split}
\delta^{\Phi} \text{S}_{\text{add}}^{(A)}= i \int \frac{d^{2|1}\zeta}{4\pi\alpha'} & ~ \frac{\alpha'}{2} \tr \left[ -A_i A_j\partial_-DX^j \right.
\\
& ~~ \left. + \left(A_j\partial_{[i}A_{k]}+A_k\partial_{[i}A_{j]}-A_i\partial_{(j}A_{k)}\right)\partial_- X^kDX^j \right]\delta^{\Phi}X^i \, .
\end{split}\label{eq:deltaSaddBasic}
\end{equation}

In the sum of \eqref{eq:deltaSanBasic} and \eqref{eq:deltaSaddBasic}, the following combination appears:
\begin{equation} \label{eq:AdelAforHP}
2\left(A_j\partial_{[i}A_{k]}+A_k\partial_{[i}A_{j]}+A_i\partial_{[k}A_{j]}\right) = -(A\text{d}A)_{ijk}+4A_k\partial_{[i}A_{j]}\, .
\end{equation}
The first term on the right hand side is minus the Chern--Simons three form in the approximation where cubic powers of the gauge field are discarded. In the full variation $\delta^{\Phi}(\text{S}+\text{S}_{\text{an}}^{(A)}+\text{S}_{\text{add}}^{(A)})$, it naturally couples to $\Gamma^+_{ijk}$ in \eqref{eq:generalCovariantVariationAction}--\eqref{eq:delX} and \eqref{eq:EoMX} and redefines its torsion to be the gauge-invariant combination:
\begin{equation} \label{eq:redefinitionCS}
\begin{split}
\delta^{\Phi}(\text{S}&+\text{S}_{\text{an}}^{(A)}+\text{S}_{\text{add}}^{(A)}) 
\\[5pt]
&= \int \frac{d^{2|1}\zeta}{4\pi\alpha'} ~ 2i \left(\Gamma+\frac{1}{2}\dd B+
\frac{\alpha'}{8}\text{CS}_3(A)\right)_{ijk}\partial_- X^jDX^k\delta^\Phi X^i + \ldots
\end{split}
\end{equation}
The complete field $H$ as in \eqref{eq:H} is generated when repeating this derivation starting with the term $\text{S}_{\text{an}}^{(\Theta)}$ also included in the effective action.

Finally, we notice that the remaining term in \eqref{eq:AdelAforHP} shares with the non-local term in \eqref{eq:deltaSanBasic} a crucial factor. Ignoring the Chern--Simons term,
\begin{equation} \label{eq:vanishingAnomaly}
\delta^\Phi (\text{S}_{\text{an}}^{(A)}+\text{S}_{\text{add}}^{(A)})=\int \frac{d^{2|1}\zeta}{4\pi\alpha'} \frac{\alpha'}{2} \tr \left( -\frac{\partial_-D}{\partial_+}\hat{A} + i\,A_k\partial_- X^k\right)2\partial_{[i}A_{j]}\delta^\Phi X^iDX^j+\ldots
\end{equation}
Keeping in mind that only quadratic terms in the gauge field are accounted for, we identify $2\partial_{[i}A_{j]}=F_{ij}$ and realise the following remarkable fact:
\vskip3pt
\textit{The term \eqref{eq:vanishingAnomaly}, which is nonlocal, is exactly cancelled if we assume the \textit{same} geometric condition \eqref{eq:Fisetcondition} that was necessary for our generalisation \eqref{eq:NonZerodelX}--\eqref{eq:NonZerodelL} of the classical Howe--Papadopoulos symmetry.}
\vskip3pt
\noindent We see this by recalling $\delta^{\Phi}X^i=\epsilon\hat{\Phi}^i$ and because \eqref{eq:vanishingAnomaly} has the factor
\begin{equation}
i_{F}(\Phi)=0\,.
\end{equation}
We knew about this constraint on $\mathcal{V}$ from the classical symmetry. Repeating the analysis with the Lorentz anomalous term $\text{S}_{\text{an}}^{(\Theta)}$ given by \eqref{eq:SanTheta}, we now obtain the same constraint on the tangent bundle $T\mathcal{M}$ as a quantum condition,
\begin{equation} \label{eq:ThetaInstanton}
i_{R^{\Theta}}(\Phi)=0\,,
\end{equation}
that is, {\it $\Theta$ must be a $\sigma$-model quasi-instanton}. Geometrically, the appearance of this condition is reasonable and in fact gives credence to our $\sigma$-model approach.  For the examples discussed in section \ref{ssec:Geometry}, where ${\cal G} = Spin(7), G_2, SU(n)$ (including both forms $(\omega,\Omega)$ in the $SU(n)$ case), the connection $\Theta$ becomes in fact a gauge-bundle instanton.  This 
extra condition is necessary for  a supersymmetric solution of a heterotic string compactification on $({\cal M}, {\cal V})$ to satisfy the supergravity equations of motion to first order in $\alpha'$ (see for example \cite{Ivanov:2009rh}). 

We conclude from this analysis that the $\mathcal{G}$-structure symmetry \eqref{eq:NonZerodelX}--\eqref{eq:NonZerodelL} is strictly-speaking anomalous. However, it can be corrected at first order in $\alpha'$ provided we impose the new target space constraint \eqref{eq:ThetaInstanton} and provided we change the torsion from $\text{d}B$ to $H$ in the classical condition,
\begin{equation}
\nabla^{+}_i\Phi_{i_1\ldots i_p}=0\,\qquad \text{(} T=H \text{)}\,.
\end{equation}
This is consistent with the redefinition induced by \eqref{eq:redefinitionCS}.

\subsubsection{Current}

Revisiting the analysis of section~\ref{sec:HPSymAnomaly}, we remark that we did not use that the infinitesimal parameter $\epsilon$ of the transformation is purely left-moving ($\partial_-\epsilon = 0$). This has implications for the $\alpha'$-corrected current associated to the symmetry (c.f. appendix~\ref{sec:SemilocalSyms}). As explained above, the variation of $\text{S}_{\text{an}}^{(A)}+\text{S}_{\text{add}}^{(A)}$ contains a term proportional to $i_F(\Phi)$ and a term which redefines the classical torsion found in $\delta^\Phi \text{S}$, eq.~\eqref{eq:generalCovariantVariationAction}--\eqref{eq:delX}. Assuming $i_F(\Phi)=0$ this means that the full $\mathcal{G}$-structure variation of $\text{S}+\text{S}^{(A)}_{\text{an}}+\text{S}^{(A)}_{\text{add}}$ (with general $\epsilon$) has exactly the same form \eqref{eq:HPdelSX} as the variation of $\text{S}$. The only difference is the redefined torsion $\dd B\rightarrow H$. We conclude that, after using all the appropriate geometric conditions,
\begin{equation}
\delta^\Phi (\text{S}+\text{S}_{\text{an}}+\text{S}_{\text{add}}) = \int \frac{\dd^2z}{4\pi\alpha'}~ \partial_-\epsilon(\zeta)\,  \dd\theta\,  (-2i)\hat \Phi\,,
\end{equation}
Therefore, remarkably, {\it the tree-level current proportional to $\hat{\Phi}$ persists at one-loop}.
Furthermore, all results thus far are true regardless of whether $\delta^\Phi_A\Lambda$ vanishes or otherwise.
The current is now conserved up to the non-local equation of motion derived from $\text{S}+\text{S}^{(A)}_{\text{an}}+\text{S}^{(A)}_{\text{add}}$, which is easy to write down from our formul\ae. This equation of motion should be interpreted as the one-loop approximation to the operator equation
\begin{equation}
\frac{\delta \Upgamma}{\delta X^i}=0\,,
\end{equation}
where $\Upgamma$ is the exact effective action. Note that the corresponding equation obtained by varying with respect to $\Lambda$ is not necessary in the conservation statement. Its role is solely to impose constraints; with no contribution to the current.

\subsubsection{Counterterms}

Effective actions are only well-defined up to finite local counterterms. These arise in particular when different schemes are used to regulate ultraviolet divergences.  In our discussion so far, we have made implicit choices when writing the effective action. We now briefly reconsider our discussion of $\mathcal{G}$-structure symmetries at order $\alpha'$ in light of these ambiguities.

The original action \eqref{eq:S_M}--\eqref{eq:S_S} is the most general covariant renormalisable $(1,0)$ supersymmetric functional. Hence, counterterms must have the same form in order not to spoil these properties \cite{Hull:1986kz,Sen:1986nm}. All the couplings in the $\sigma$-model (metric, $B$-field, gauge field and $S$) have corresponding counterterms
$\Delta G_{ij}$, $ \Delta B_{ij}$, $\Delta A_i{}^\alpha{}_\beta$, and $\Delta S_\alpha$. We define $\widetilde{G}_{ij}=G_{ij}+\Delta G_{ij}$, and similarly for the others, and add tildes to identify quantities constructed from such redefined tensors. In this section, we also write explicitly the dependence of action functionals on target space tensors: for example, the allowed counterterms are collectively written as $\text{S}_{\text{c.t.}}=\text{S}(\Delta G,\Delta B,\Delta A,\Delta S)$. The one-loop effective action, with counterterms, is then taken as
\begin{equation} \label{eq:CTEffAction}
\text{S}(\widetilde{G},\widetilde{B},\widetilde{A},\widetilde{S})+\text{S}^{(A)}_{\text{an}}(A)+\text{S}^{(A)}_{\text{add}}(A)\, ,
\end{equation}
where we still ignore $\text{S}_{\text{an}}^{(\Theta)}$ to simplify the discussion.

It is important to distinguish carefully the gauge fields in \eqref{eq:CTEffAction} from the gauge field in the symmetry variation. We must use the same symmetry as before, namely \eqref{eq:NonZerodelX}--\eqref{eq:NonZerodelL},
\begin{align}
&\delta^\Phi X^i = \epsilon\hat{\Phi}^i\,,
&\delta^\Phi\Lambda^\alpha + A_i{}^\alpha{}_\beta\Lambda^\beta\delta^\Phi X^i = \epsilon\hat{\Upsilon}^\alpha{}_\beta\,\frac{\Delta \text{S}(A,S)}{\Delta \Lambda_\beta}\,, 
\end{align}
but computations are simpler if we write this variation of $\Lambda$ as
\begin{equation}
\delta^\Phi_{\widetilde{A}}\Lambda^\alpha = (\Delta A)_i{}^\alpha{}_\beta\Lambda^\beta\delta^\Phi X^i+\epsilon\hat{\Upsilon}^\alpha{}_\beta\,\frac{\Delta \text{S}(A,S)}{\Delta \Lambda_\beta}\,. \label{eq:FunnyDeltaLambda}
\end{equation}

Focusing for the moment on the symmetry variation of $\text{S}(\widetilde{G},\widetilde{B},\widetilde{A},\widetilde{S})$ with respect to $X$, we find equation~\eqref{eq:HPdelSX} again, this time written in terms of tilde tensors
\begin{equation}
\begin{split}
\delta \text{S} &= \int \frac{\dd^2z}{4\pi\alpha'}\, \epsilon(\zeta) \dd\theta
\bigg[
-\frac{2i}{p} \partial_- X^i DX^{j_1} \widetilde{\nabla}_i^+(\widetilde{G}_{jj_1}\hat{\Phi}^j)
\\[3pt]
&\qquad\qquad \qquad \qquad \qquad +   \tr\Big(
\Lambda \widetilde{F}_{ij}\, D X^j\hat \Phi^i \Lambda
+ (-1)^{p-1} m\, (\dd_{\widetilde{A}} \widetilde{S})_i\,\hat \Phi^i \Lambda\Big)
\bigg]
\\[3pt]
&\quad
+ \  \int_\Sigma \frac{\dd^2z}{4\pi\alpha'}~ \partial_-\epsilon(\zeta)\,  \dd\theta\,  (-2i)\frac{1}{p!}\widetilde{G}_{jj_1}\Phi^j{}_{j_2\ldots j_p}DX^{j_1\ldots j_p}\,. 
\end{split}\label{eq:CTeasy}
\end{equation}

Meanwhile $\text{S}^{(A)}_{\text{an}}+\text{S}^{(A)}_{\text{add}}$ is independent of $\Lambda$ and its variation with respect to $X$ is exactly as in section~\ref{sec:HPSymAnomaly}. It produces the non-local term \eqref{eq:vanishingAnomaly} and a term which redefines the torsion in \eqref{eq:CTeasy} to be
\begin{equation}
T = \text{d}\widetilde{B}+ \frac{\alpha'}{4}\text{CS}_3(A)\,.
\end{equation}
as in \eqref{eq:redefinitionCS}.

Finally, we account for the variation of $\text{S}(\widetilde{G},\widetilde{B},\widetilde{A},\widetilde{S})$ due to \eqref{eq:FunnyDeltaLambda}. We find
\begin{equation}
\begin{split}
\delta\text{S} &= \int \frac{\dd^{2|1}\zeta}{4\pi\alpha'} \bigg[\frac{\Delta \text{S}(A,S)}{\Delta\Lambda_\alpha}\bigg((\Delta A)_{i\alpha\beta}\Lambda^\beta\delta^\Phi X^i - \big(2\,\widehat{\Delta A}^\beta{}_\gamma\Lambda^\gamma +  m\, \Delta S^\beta\big)\epsilon\hat{\Upsilon}_{\alpha\beta} \bigg) \\[2pt]
&\qquad\qquad\qquad\qquad\qquad\qquad\qquad + m\, \Delta S^\alpha(\Delta A)_{i\alpha\beta}\Lambda^\beta\delta^\Phi X^i \bigg]\,.
\end{split}\label{eq:CTcomplicated}
\end{equation}
To obtain this we used $\Upsilon_{(\alpha\beta)}=0$ and
\begin{equation}
\frac{\Delta \text{S}(\widetilde{A},\widetilde{S})}{\Delta\Lambda_\alpha} = 2\, D_{\widetilde{A}}\Lambda^\alpha 
+  m\, \widetilde{S}^\alpha = \frac{\Delta \text{S}(A,S)}{\Delta\Lambda_\alpha} + 2\,\widehat{\Delta A}^\alpha{}_\beta\Lambda^\beta + m\, \Delta S^\alpha \,.
\end{equation}

The next step is to group the terms in the full variation (the sum of \eqref{eq:CTeasy}, \eqref{eq:vanishingAnomaly}, and \eqref{eq:CTcomplicated}) sharing the same powers of the fundamental superfields and their derivatives. From their prefactors, it is straightforward to read off constraints on counterterms and target space tensors ensuring preservation of the symmetry. We leave the general case to the reader, and focus here on the most commonly encountered counterterms $\Delta G$ and $\Delta B$.

If we set $\Delta A = 0$ and $\Delta S = 0$, then \eqref{eq:CTcomplicated} does not interfere with \eqref{eq:CTeasy} and we can read off, much like before, the condition
\begin{equation}
i_F(\Phi)=0\,,
\end{equation}
from the non-local term and, from \eqref{eq:CTeasy},
\begin{align}
&\widetilde{\nabla}_i^+(\widetilde{G}_{j[j_1}\Phi^j{}_{j_2\ldots j_p]})=0 \qquad 
\left( T = \text{d}\widetilde{B}+ \frac{\alpha'}{4}\text{CS}_3(A) \right) \,, \label{eq:CTcondition}\\ 
&i_{\dd_A S}(\Phi)=0\,,
\end{align}
and the current
\begin{equation}
2i(-1)^p\frac{1}{p!}\widetilde{G}_{jj_1}\Phi^j{}_{j_2\ldots j_p}DX^{j_1\ldots j_p} = 2i(-1)^p\left( \hat{\Phi} + \frac{1}{p}\Delta G_{ij}DX^{i}\hat{\Phi}^j \right) \,.
\end{equation}

\subsection{Superconformal anomalies} \label{sec:Superconformal anomalies}

Appendix~\ref{sec:SemilocalSyms} gives a short account of the basic features of $(1,0)$ superconformal symmetry in our non-linear $\sigma$-model (for vanishing mass). It is tantalizing to try on this symmetry the anomaly analysis presented in the last section, using the effective action. A good motivation to treat all symmetries on the same footing is in prevision to study the algebra they form. This is particularly interesting at the quantum level. In the case of superconformal symmetries, we have a prejudice on the outcome based on the substantial literature on conformal anomalies in two dimensional $\sigma$-models (see e.g.\ \cite{Hull:1987yi, Hull:1985rc, Sen:1985qt, Lambert:1995hs, Hull:1986hn, Hull:1986kz, Hull:1985zy, Hull:1987pc, Hull:1987yi, Callan:1989nz, AlvarezGaume:1983ab}). Nevertheless the method we use, based on the effective action \eqref{eq:SlavnovTaylor}, is non-standard in this context. As a complement to our discussion of $\mathcal{G}$-structure anomalies, it is worthwhile to connect our angle of analysis with classical string theory lore.

Our main result is simply that the superconformal variation of the one-loop effective action \eqref{eq:SanTheta} vanishes,
\begin{equation} \label{eq:deltaSan=0(conf)}
\delta^\epsilon \text{S}_{\text{an}}^{(A)} = 0\,.
\end{equation}
This fact will be proven shortly. It follows after some algebraic manipulations only, without using any equations of motion and without imposing any constraints on the $\sigma$-model couplings.

Naively, the conclusion is that superconformal symmetries are not anomalous at one-loop, which is consistent with the expectation that a nearby superconformal fixed point exists in the universality class of $\text{S}$. However, this should only be true for certain configurations of the $\sigma$-model couplings: those which satisfy effective target space equations of motion \cite{Callan:1985ia,Hull:1986kz}.

To reconcile \eqref{eq:deltaSan=0(conf)} with the literature, it is useful to reconsider the calculation of the effective action itself. Along the way, ultraviolet divergences are generated and are renormalised away in redefined couplings \cite{Friedan:1980jf,Howe:1986vm}. This generates beta functionals for the metric, B-field and gauge field, which must be trivial (not necessarily zero) to guarantee scale invariance. It is at this step that the familiar constraints on the couplings arise. Only for those configurations satisfying the target space equations of motion is the model scale invariant.

After renormalisation, there remains in the effective action ultraviolet-finite terms only, which are all expressed in terms of renormalised quantities. The term $\text{S}_{\text{an}}^{(A)}$ that we have been using and the whole discussion of this section, were in terms of renormalised objects. At this level, the fact that we find $\delta \text{S}_{\text{an}}^{(A)}=0$, and thus no further restrictions by imposing conformal symmetry, can essentially\footnote{Strictly speaking it is best to revisit \cite{Mavromatos:1988jj} Zamolodchikov's theorem when working with non-linear $\sigma$-models. Assumptions sometimes fail, such as discreteness of the spectrum for noncompact target manifolds and unitarity for Lorentzian signature \cite{Polchinski:1987dy}.} be understood from the argument that scale invariant theories in two dimensions are automatically conformal \cite{Zamolodchikov:1986gt}.

We now prove \eqref{eq:deltaSan=0(conf)}. It is useful to break the proof into two steps. First we show that the superconformal variation is local. Then we show that it vanishes. The derivation starts as in section~\ref{sec:HPSymAnomaly} and we can reuse \eqref{eq:deltaSanBasic},
\begin{equation}
\delta^\Phi \text{S}_{\text{an}}^{(A)}= \int \frac{d^{2|1}\zeta}{4\pi\alpha'} ~ \frac{\alpha'}{2} \text{tr} \left[-2\left(\frac{\partial_-D}{\partial_+}\hat{A}\right)\partial_{[i} A_{j]}DX^j+i(\partial_- \hat{A}) A_i\right]\delta^{\epsilon}X^i.
\end{equation} 
now for superconformal transformations \eqref{eq:susy1}
\begin{equation}
\delta^\epsilon X^i = i \epsilon \partial_+ X^i + \frac{1}{2} D \epsilon D X^i\,.
\end{equation}
Focusing on the non-local part, we notice that
\begin{equation}
2\partial_{[i}A_{j]}DX^j\delta^\epsilon X^i = D(\epsilon DA_iDX^i)\,.
\end{equation}
Integrating by parts with $D$, we find the local representation
\begin{equation} \label{eq:deltaSanConfIntermediate}
\delta^\epsilon \text{S}_{\text{an}}^{(A)}=i \int \frac{d^{2|1}\zeta}{4\pi\alpha'} ~ \frac{\alpha'}{2}\,\text{tr}\left[\partial_-\hat{A}(\epsilon DA_iDX^i+A_i\delta^\epsilon X^i)\right].
\end{equation}

We now show that this vanishes. It is useful to define the operator
\begin{equation}
D_\epsilon = \epsilon D+\frac{1}{2}D\epsilon
\end{equation}
so that $\delta^\epsilon X^i = D_\epsilon DX^i$ and identify in~\eqref{eq:deltaSanConfIntermediate}
\begin{equation}
\epsilon DA_iDX^i+A_i\delta^\epsilon X^i = D_\epsilon \hat{A}\,.
\end{equation}
Then, integrating by parts with $\partial_-$,
\begin{equation} \label{eq:deltaSanConf-}
\delta^\epsilon \text{S}_{\text{an}}^{(A)}=-i\int \frac{d^{2|1}\zeta}{4\pi\alpha'} ~ \frac{\alpha'}{2}\,\text{tr}(\hat{A}\partial_-D_\epsilon\hat{A})\,.
\end{equation}
Alternatively, we can integrate by parts with $D_\epsilon$. Indeed, it is easy to prove that, given two superfields $U$ and $V$
\be
(D_\epsilon U) V + (-1)^F\, UD_\epsilon V = D(\epsilon\, UV)\, ,
\ee
where $F= +1$ if $U$ is a commuting superfield and $F=-1$ if it is anticommuting.
From \eqref{eq:deltaSanConfIntermediate} this yields
\begin{equation} \label{eq:deltaSanConf+}
\delta^\epsilon \text{S}_{\text{an}}^{(A)}=i \int \frac{d^{2|1}\zeta}{4\pi\alpha'} ~ \frac{\alpha'}{2}\,\text{tr}\big((D_\epsilon\partial_-\hat{A})\hat{A}\big)
=i\int \frac{d^{2|1}\zeta}{4\pi\alpha'} ~ \frac{\alpha'}{2}\,\text{tr}(\hat{A}D_\epsilon\partial_-\hat{A})\,,
\end{equation}
where we have used cyclicity of the trace in the last step. We complete the proof of \eqref{eq:deltaSan=0(conf)} by comparing \eqref{eq:deltaSanConf+} and \eqref{eq:deltaSanConf-}, and by using $[\partial_-,D_\epsilon]=0$, which follows from $\partial_-\epsilon = 0$.

For a general symmetry parameter, we have instead
\begin{equation} \label{eq:deltaSanConfFinal}
\delta^\epsilon \text{S}_{\text{an}}^{(A)}=i\int \frac{d^{2|1}\zeta}{4\pi\alpha'} ~ \frac{\alpha'}{4}\,\text{tr}(\hat{A}[D_\epsilon,\partial_-]\hat{A})
=-i\int \frac{d^{2|1}\zeta}{4\pi\alpha'} ~ \frac{\alpha'}{4}\partial_-\epsilon \, \text{tr}(\hat{A}D\hat{A})\,.
\end{equation}

\subsubsection{Gauge-invariant supercurrent at order $\alpha'$}

As an application of the proof above, we now derive the $\alpha'$-correction to the left-moving stress-tensor and supersymmetry currents of generic massless $(1,0)$ $\sigma$-models \eqref{eq:S_M}--\eqref{eq:S_V}. To the best of our knowledge, this calculation is new. Classically, the Noether procedure yields the superfield \eqref{eq:T+}
\be \label{eq:T+text}
{\cal T}_{\neq+} =  G_{ij}\, \partial_+ X^i D X^j 
- i\, \widehat{\dd B}\, .
\ee
This supercurrent is right-moving on shell, $\partial_-\mathcal{T}_{\neq+}\approx 0$, and is composed of the supersymmetry current $G_{\neq+}$ and stress-tensor $T_{\neq\neq}$. The second term in \eqref{eq:T+text} is often discarded in the literature. At order $\alpha'$, $\dd B$ is not gauge-invariant, as reviewed in section~\ref{sec:GreenSchwarz}. It is natural to ask if our considerations from this section can fix this issue.

It turns out they do. To see this, we extract from \eqref{eq:deltaSanConfFinal} the contribution of $\text{S}_{\text{an}}^{(A)}$ to $\mathcal{T}_{\neq+}$,
\begin{equation}
- i\,\frac{\alpha'}{4}\,\tr(\hat{A}D\hat{A})\,.
\end{equation}
Substituting $D\hat{A}=\hat{F}+ iA_i\partial_+X^i$
this is composed of two terms. The first one immediately yields the Chern--Simons correction necessary to make $\mathcal{T}_{\neq+}$ gauge-invariant:
\begin{equation}
\tr(\hat{A}\hat{F})=\widehat{\text{CS}_3(A)}\,,
\end{equation}
up to corrections cubic in $\hat{A}$. The second term can be absorbed by the variation of $\text{S}_{\text{add}}^{(A)}$. A particularly easy way to see this is to remember that $\text{S}_{\text{add}}^{(A)}$ is a metric counterterm, so we can read off its contribution to the current directly from \eqref{eq:T+text}. With $\Delta G_{ij}= - \frac{\alpha'}{4}\,\tr(A_i A_j)$, this is
\begin{equation}
\Delta G_{ij}\partial_+ X^i DX^j = - \frac{\alpha'}{4}\,\tr(A_i A_j)\partial_+ X^i DX^j\,.
\end{equation}

More generally, the impact of changing counterterms is easy to analyse for superconformal transformations. Assuming counterterms of the form of the classical action, with $G$ replaced by $\Delta G$, and similarly for the other couplings, superconformal invariance cannot be spoiled. Indeed, no assumption on the couplings are made to prove classical superconformal invariance. As for the current, the modifications are as discussed in the case of $\text{S}_{\text{add}}^{(A)}$. The most general form of the $\alpha'$-corrected supercurrent, including counterterms, is
\be
{\cal T}_{\neq+} =  (G_{ij}+\Delta G_{ij})\, \partial_+ X^i D X^j 
-i (\widehat{H} + \widehat{\dd (\Delta B)})\, .
\ee

\subsection{A caveat: gauge-invariant contributions to $\Upgamma$} \label{sec:Caveat}

It should be stressed that our analysis of $\alpha'$-corrections in this section has turned out to be much simpler than it should perhaps have been. There is an important caveat to our analysis, which we now point out even if it seems to be largely unimportant given the sensible results obtained so far in section~\ref{sec:Anomalies}.

As they were primarily interested in the Green--Schwarz mechanism, the authors of \cite{Hull:1986xn} focused only on Yang--Mills and Lorentz non-covariance in the $\sigma$-model one-loop effective action, leading to what we called $\text{S}_{\text{an}}$. Analyses of gauge anomalies at higher loops have been performed \cite{Hamada:1987ph,Grignani:1987gk,Foakes:1988wy,Ellwanger:1988cc,Lambert:1995hs}. However, we have not been able to locate in the existing literature a more complete calculation of the effective action which would include all covariant terms.\footnote{One particular covariant but infrared divergent term was reported in \cite{Hull:1986xn}. We have not included it in our present analysis given that further terms on the same footing are expected to exist, and should be analysed together.} Such terms are crucial to our analysis because they may lead to anomalies of $\mathcal{G}$-structure (and superconformal) symmetries even if they do not produce gauge and gravity anomalies.

The fact that our results at order $\alpha'$ so far nicely align with supergravity expectations suggests that this problem in fact does not arise. More precisely, we conjecture that gauge and Lorentz invariant contributions to the effective action are automatically invariant under $\mathcal{G}$-structure symmetries --- up to the usual target space conditions \eqref{eq:HPcondition}--\eqref{eq:Mcondition}.
We hope to report on this conjecture more fully in a future communication.

\section{Conclusion}

Our main result in this work is the generalisation \eqref{eq:NonZerodelX}--\eqref{eq:NonZerodelL} of the symmetry of \cite{Howe:1991im} holding for general $(1,0)$ non-linear $\sigma$-models with non-Abelian background gauge fields turned on and also possibly a mass term. This symmetry is defined with a target space $p$-form $\Phi$ as well as a tensor $\Upsilon\in \Omega^{p-2}(\mathcal{M},\bigwedge^2\mathcal{V})$, which may, or may not, be chosen to vanish identically. The constraints \eqref{eq:AllConditions} on these tensors and the couplings of the $\sigma$-model are strongly reminescent of the supersymmetry conditions appearing in the context of heterotic compactifications. In fact, for the cases of $Spin(7)$ and $G_2$ compactifications that we discussed more closely, these conditions are equivalent. Contrastingly in the $SU(3)$ case, our symmetry does not require an integrable complex structure, but this can be enforced by demanding that it generates with $(1,0)$ superconformal symmetry the $(2,0)$ algebra.

We have demonstrated moreover how a modified version of our $\mathcal{G}$-structure symmetry persists quantum-mechanically. There remains caveats to this statement: crucially, a complete calculation of the $\sigma$-model one-loop effective action at first order in $\alpha'$ is necessary for definitive conclusions. Nevertheless, our analysis based only on the non-local term $S_{\text{an}}$ has already produced quantum conditions impressively close to the supergravity expectations, such as the quasi-instanton condition $i_{R^{\Theta}}(\Phi)=0$ on $T\mathcal{M}$.

The conserved current for all the $\mathcal{G}$-structure symmetries that we considered is the operator $\hat{\Phi}$ naturally associated to the differential $p$-form. This remains true when including $\alpha'$-corrections but can be affected by metric counterterms.

Superconformal transformations were also discussed from the angle of the quantum effective action and compared with string theory. Our results at order $\alpha'$ are all consistent with Green--Schwarz gauge-invariance and the heterotic Bianchi identity.

The most immediate application of our $\mathcal{G}$-structure symmetry is in finding marginal deformations of $\sigma$-models used as internal sectors in heterotic string compactifications. This project was started by the authors in \cite{Melnikov:2011ez,Fiset:2017auc} for the case where $\alpha'=0$.  
By isolating explicitly the symmetry associated with supersymmetric backgrounds, it becomes clear how to impose that it be preserved by deformations.
One consequence of this study is the better understanding, to first order in $\alpha'$, of the relation between nilpotent operators which describe marginal deformations in terms of their cohomology and analogous operators formulated in supergravity \cite{delaOssa:2014cia,Anderson:2014xha,Garcia-Fernandez:2015hja,Clarke:2016qtg,delaOssa:2016ivz,delaOssa:2017pqy,Ashmore:2018ybe,Garcia-Fernandez:2018ypt}. This will be the subject of a forthcoming publication.

Besides, it is likely that there will be some connections of our results with the considerations of \cite{Ekstrand:2009zd,Ekstrand:2010wu}, where the Chiral de Rham complex \cite{Malikov:1998dw} was likened to a formal quantisation of the $(1,1)$ nonlinear $\sigma$-model. In these papers, $\Lambda$-brackets were proposed as a way to interpolate between special holonomy OPE algebras \cite{Odake:1988bh,Shatashvili:1994zw} and the classical symmetries of \cite{Howe:1991im}. A more detailed comprehension of commutator and current algebras of our extended $\mathcal{G}$-structure symmetries would make a useful start about this. This is especially interesting at order $\alpha'$, where the condition $\dd H = 0$ fails, suggesting radical alterations to the algebras. We hope to return to these issues in the near future.
 
It will also be interesting to clarify if our symmetry perhaps can be thought of as the infrared limit of some useful symmetry of gauged linear $\sigma$-model (see e.g. \cite{McOrist:2010ae} and references therein).

More speculatively, since $N=2$ supersymmetry is a subcase of $\mathcal{G}$-structure symmetries, it is permitted to think that some of the powerful tools following from the former admit a non-linear generalisation to the latter. We might ask for example for a ``$\mathcal{G}$-structure'' analogue of supersymmetric localisation, to name but one, which would encompass $(2,0)$ localisation \cite{Closset:2015ohf}. In any case, whenever they are preserved, these symmetries put strong constraints on the dynamics of the $\sigma$-model and should guide the study of string vacua in the $\alpha'$ expansion from a world-sheet point of view \cite{Gross:1986iv, Grisaru:1986px, Candelas:1986tz}. They might find applications for instance to generalise the results of \cite{Nemeschansky:1986yx} to target spaces other than Calabi--Yau manifolds \cite{Jardine:2018sft,Becker:2014rea}.

\section*{Acknowledgement}

We thank Eirik Eik Svanes who asked the question leading to this research. We are also grateful for the generous comments of Chris Hull and Callum Quigley. MAF also wishes to acknowledge useful discussions with Pietro Benetti Genolini and Pyry Kuusela. XD would like to thank the Korea Institute for Advanced Study for hospitality, where some of this work was completed. The work of XD is supported in part by the EPSRC grant EP/J010790/1.
The research of MAF is financed by a Reidler scholarship from the Mathematical Institute at the University of Oxford and by a FRQNT doctoral scholarship from the Government of Quebec.

\appendix

\section{Semi-local and superconformal symmetries} \label{sec:SemilocalSyms}

\subsection{Semi-local symmetries}

We give here a brief account of continuous semi-local (or chiral) symmetries of the action \eqref{eq:S_M}--\eqref{eq:S_S}, of which the $\cal G$-structure symmetry is an example. Such symmetries sit somewhere between global and local symmetries in that they are parametrized, in their infinitesimal version, by a small function $\epsilon(\zeta)$ depending on some, but not all, of the superspace coordinates. Equivalently, $\epsilon(\zeta)$ is a constrained parameter. It could be Grassmann even or odd. A symmetry transformation is generally given as
\begin{align}
\delta X^i &= \overline{\delta X^i}(\epsilon,X,\Lambda) \, , \label{eq:BarSymmetry1}\\
\delta \Lambda^\alpha &= \overline{\delta \Lambda^\alpha}(\epsilon,X,\Lambda)\, , \label{eq:BarSymmetry2}
\end{align}
where the right hand sides are specific expressions involving the infinitesimal parameter, the fundamental fields and, in general, their (super)derivatives. The statement of symmetry is that the induced variation of the action can be recasted in the form
\begin{equation} \label{eq:Noether}
\overline{\delta \text{S}} 
= \int \frac{d^{2|1}\zeta}{4\pi\alpha'} \left(\frac{\delta \text{S}}{\delta X^i}\, \overline{\delta X^i} 
+ \frac{\delta \text{S}}{\delta \Lambda^\alpha}\, \overline{\delta \Lambda^\alpha}\right) = \int \frac{d^{2|1}\zeta}{4\pi\alpha'} ~ \partial_\mu \epsilon(\zeta) J^\mu\, ,
\end{equation}
where we must still define the right hand side. The first equality here is general for arbitrary variations, while the second is specific to the barred symmetry variations. The index $\mu$ covers all directions $(z^+,z^-,\theta)$ of superspace, but in the case of semi-local symmetries, at least one of the superfields $J^+$, $J^-$, $J^\theta$ vanishes identically. Because of this, $\overline{\delta S}=0$ to leading order if we impose the constraint $\partial_{\bar{\mu}} \epsilon(\zeta)=0$ for all directions $\bar{\mu}$ with non-vanishing $J^{\bar{\mu}}$.

Integrating by parts, \eqref{eq:Noether} is Noether's theorem in $(1,0)$ superspace. When $\epsilon(\zeta)$ is freed from its constraint, i.e.\ made fully local on superspace, then $\overline{\delta S}\neq 0$, but the second equality in \eqref{eq:Noether} still holds. The familiar local current conservation rule $\partial_{\bar{\mu}}J^{\bar{\mu}}\approx 0$  then follows from the fact that $\epsilon(\zeta)$ is made to depend on \textit{all} integration variables (including $\theta$), and from the equations of motion. We use curly equal signs for equations holding on-shell.

\subsection{Superconformal symmetry}

Arguably the most important examples of chiral symmetries are conformal transformations. We focus on symmetries acting on the supersymmetric side ($+$) of the massless $\sigma$-model \eqref{eq:S_M}--\eqref{eq:S_V}. Consider the transformation 
\begin{align}
\delta^\epsilon X^i &= i \epsilon \partial_+ X^i + \frac{1}{2} D \epsilon D X^i\, ,\label{eq:susy1}
\\
\delta^\epsilon_A \Lambda^\alpha &= i\epsilon \partial_{+A} \Lambda^\alpha 
+ \frac{1}{2} D \epsilon D_A \Lambda^\alpha\, ,\label{eq:susy2}
\end{align}
where $\epsilon(\zeta)$ is an infinitesimal function of the world-sheet coordinates and
\be
D_A \Lambda = D \Lambda +  \hat{A} \Lambda\, , \qquad
\partial_{+A} \Lambda = \partial_+ \Lambda + (A_i\,\partial_+ X^i)\Lambda\, .
\ee
The statement in this case is that the massless action is invariant under these superconformal  transformations 
whenever  $\epsilon=\epsilon(z^+, \theta)$. The chiral supercurrent associated to this symmetry is given by 
\be \label{eq:T+}
{\cal T}_{\neq +} =  G_{ij}\, \partial_+ X^i D X^j 
-i\widehat{\dd B}\, .
\ee
Note that this is the same current as the one obtained when $\Lambda = 0$.

\newpage

\bibliographystyle{alpha}

\bibliography{FisetNOV2018}

\newcommand{\etalchar}[1]{$^{#1}$}
\begin{thebibliography}{ADLOM{\etalchar{+}}18}

\bibitem[Ach98]{Acharya:1997rh}
Bobby~Samir Acharya.
\newblock {On mirror symmetry for manifolds of exceptional holonomy}.
\newblock {\em Nucl. Phys.}, B524:269--282, 1998.

\bibitem[ADLOM{\etalchar{+}}18]{Ashmore:2018ybe}
Anthony Ashmore, Xenia De~La~Ossa, Ruben Minasian, Charles
  Strickland-Constable, and Eirik~Eik Svanes.
\newblock {Finite deformations from a heterotic superpotential: holomorphic
  Chern--Simons and an $L_\infty$ algebra}.
\newblock 2018.

\bibitem[AGF81]{AlvarezGaume:1981hm}
Luis Alvarez-Gaume and Daniel~Z. Freedman.
\newblock {Geometrical Structure and Ultraviolet Finiteness in the
  Supersymmetric Sigma Model}.
\newblock {\em Commun. Math. Phys.}, 80:443, 1981.

\bibitem[AGF83]{AlvarezGaume:1983ab}
Luis Alvarez-Gaume and Daniel~Z. Freedman.
\newblock {Potentials for the Supersymmetric Nonlinear Sigma Model}.
\newblock {\em Commun. Math. Phys.}, 91:87, 1983.

\bibitem[AGFM81]{AlvarezGaume:1981hn}
Luis Alvarez-Gaume, Daniel~Z. Freedman, and Sunil Mukhi.
\newblock {The Background Field Method and the Ultraviolet Structure of the
  Supersymmetric Nonlinear Sigma Model}.
\newblock {\em Annals Phys.}, 134:85, 1981.

\bibitem[AGG85]{AlvarezGaume:1985yb}
Luis Alvarez-Gaume and Paul~H. Ginsparg.
\newblock {GEOMETRY ANOMALIES}.
\newblock {\em Nucl. Phys.}, B262:439--462, 1985.

\bibitem[AGS14]{Anderson:2014xha}
Lara~B. Anderson, James Gray, and Eric Sharpe.
\newblock {Algebroids, Heterotic Moduli Spaces and the Strominger System}.
\newblock {\em JHEP}, 07:037, 2014.

\bibitem[BCZ85]{Braaten:1985is}
Eric Braaten, Thomas~L. Curtright, and Cosmas~K. Zachos.
\newblock {Torsion and Geometrostasis in Nonlinear Sigma Models}.
\newblock {\em Nucl. Phys.}, B260:630, 1985.
\newblock [Erratum: Nucl. Phys.B266,748(1986)].

\bibitem[BDFM88]{Banks:1987cy}
Tom Banks, Lance~J. Dixon, Daniel Friedan, and Emil~J. Martinec.
\newblock {Phenomenology and Conformal Field Theory Or Can String Theory
  Predict the Weak Mixing Angle?}
\newblock {\em Nucl. Phys.}, B299:613--626, 1988.

\bibitem[BDZ17]{Braun:2017ryx}
Andreas~P. Braun and Michele Del~Zotto.
\newblock {Mirror Symmetry for $G_2$-Manifolds: Twisted Connected Sums and Dual
  Tops}.
\newblock {\em JHEP}, 05:080, 2017.

\bibitem[BDZ18]{Braun:2017csz}
Andreas~P. Braun and Michele Del~Zotto.
\newblock {Towards Generalized Mirror Symmetry for Twisted Connected Sum $G_2$
  Manifolds}.
\newblock {\em JHEP}, 03:082, 2018.

\bibitem[Bis89]{MR1006380}
Jean-Michel Bismut.
\newblock A local index theorem for non-{K}\"{a}hler manifolds.
\newblock {\em Math. Ann.}, 284(4):681--699, 1989.

\bibitem[BNY85]{Bagger:1985pw}
Jonathan Bagger, Dennis Nemeschansky, and S.~Yankielowicz.
\newblock {ANOMALY CONSTRAINTS ON NONLINEAR SIGMA MODELS}.
\newblock {\em Nucl. Phys.}, B262:478, 1985.
\newblock [,478(1985)].

\bibitem[BRW14]{Becker:2014rea}
Katrin Becker, Daniel Robbins, and Edward Witten.
\newblock {The $\alpha'$ Expansion On A Compact Manifold Of Exceptional
  Holonomy}.
\newblock {\em JHEP}, 06:051, 2014.

\bibitem[Bry05]{Bryant:2005mz}
Robert~L. Bryant.
\newblock {Some remarks on G(2)-structures}.
\newblock 2005.

\bibitem[BSW97]{Blumenhagen:1996vu}
Ralph Blumenhagen, Rolf Schimmrigk, and Andreas Wisskirchen.
\newblock {(0,2) mirror symmetry}.
\newblock {\em Nucl. Phys.}, B486:598--628, 1997.

\bibitem[CFP{\etalchar{+}}86]{Candelas:1986tz}
P.~Candelas, M.~D. Freeman, C.~N. Pope, M.~F. Sohnius, and K.~S. Stelle.
\newblock {Higher Order Corrections to Supersymmetry and Compactifications of
  the Heterotic String}.
\newblock {\em Phys. Lett.}, B177:341--346, 1986.

\bibitem[CGFT16]{Clarke:2016qtg}
Andrew Clarke, Mario Garcia-Fernandez, and Carl Tipler.
\newblock {Moduli of $G_2$ structures and the Strominger system in dimension
  7}.
\newblock 2016.

\bibitem[CGJS16]{Closset:2015ohf}
Cyril Closset, Wei Gu, Bei Jia, and Eric Sharpe.
\newblock {Localization of twisted $ \mathcal{N}=\left(0,\;2\right) $ gauged
  linear sigma models in two dimensions}.
\newblock {\em JHEP}, 03:070, 2016.

\bibitem[CMPF85]{Callan:1985ia}
Curtis~G. Callan, Jr., E.~J. Martinec, M.~J. Perry, and D.~Friedan.
\newblock {Strings in Background Fields}.
\newblock {\em Nucl. Phys.}, B262:593--609, 1985.

\bibitem[CT89]{Callan:1989nz}
Curtis~G. Callan, Jr. and Larus Thorlacius.
\newblock {SIGMA MODELS AND STRING THEORY}.
\newblock In {\em {Theoretical Advanced Study Institute in Elementary Particle
  Physics: Particles, Strings and Supernovae (TASI 88) Providence, Rhode
  Island, June 5-July 1, 1988}}, pages 795--878, 1989.

\bibitem[DeW67]{DeWitt:1967uc}
Bryce~S. DeWitt.
\newblock {Quantum Theory of Gravity. 3. Applications of the Covariant Theory}.
\newblock {\em Phys. Rev.}, 162:1239--1256, 1967.
\newblock [,307(1967)].

\bibitem[dlOLS16]{delaOssa:2016ivz}
Xenia de~la Ossa, Magdalena Larfors, and Eirik~Eik Svanes.
\newblock {Infinitesimal moduli of G2 holonomy manifolds with instanton
  bundles}.
\newblock {\em JHEP}, 11:016, 2016.

\bibitem[dlOLS18]{delaOssa:2017pqy}
Xenia de~la Ossa, Magdalena Larfors, and Eirik~E. Svanes.
\newblock {The Infinitesimal Moduli Space of Heterotic G$_{2}$ Systems}.
\newblock {\em Commun. Math. Phys.}, 360(2):727--775, 2018.

\bibitem[dlOS14]{delaOssa:2014cia}
Xenia de~la Ossa and Eirik~E. Svanes.
\newblock {Holomorphic Bundles and the Moduli Space of N=1 Supersymmetric
  Heterotic Compactifications}.
\newblock {\em JHEP}, 10:123, 2014.

\bibitem[EFS89]{Ellwanger:1988cc}
U.~Ellwanger, J.~Fuchs, and M.~G. Schmidt.
\newblock {The Heterotic $\sigma$ Model With Background Gauge Fields}.
\newblock {\em Nucl. Phys.}, B314:175, 1989.

\bibitem[EHKZ09]{Ekstrand:2009zd}
Joel Ekstrand, Reimundo Heluani, Johan Kallen, and Maxim Zabzine.
\newblock {Non-linear sigma models via the chiral de Rham complex}.
\newblock {\em Adv. Theor. Math. Phys.}, 13(4):1221--1254, 2009.

\bibitem[EHKZ13]{Ekstrand:2010wu}
Joel Ekstrand, Reimundo Heluani, Johan Kallen, and Maxim Zabzine.
\newblock {Chiral de Rham complex on Riemannian manifolds and special
  holonomy}.
\newblock {\em Commun. Math. Phys.}, 318:575--613, 2013.

\bibitem[FI02]{Friedrich:2001nh}
Thomas Friedrich and Stefan Ivanov.
\newblock {Parallel spinors and connections with skew symmetric torsion in
  string theory}.
\newblock {\em Asian J. Math.}, 6:303--336, 2002.

\bibitem[FI03]{Friedrich:2001yp}
Thomas Friedrich and Stefan Ivanov.
\newblock {Killing spinor equations in dimension 7 and geometry of integrable
  G(2) manifolds}.
\newblock {\em J. Geom. Phys.}, 48:1, 2003.

\bibitem[Fis18]{Fiset:2018huv}
Marc-Antoine Fiset.
\newblock {Superconformal algebras for twisted connected sums and $G_2$ mirror
  symmetry}.
\newblock 2018.

\bibitem[FMR88]{Foakes:1988wy}
A.~P. Foakes, N.~Mohammedi, and D.~A. Ross.
\newblock {Three Loop Beta Functions for the Superstring and Heterotic String}.
\newblock {\em Nucl. Phys.}, B310:335--354, 1988.

\bibitem[FO97]{Figueroa-OFarrill:1996tnk}
Jose~M. Figueroa-O'Farrill.
\newblock {A Note on the extended superconformal algebras associated with
  manifolds of exceptional holonomy}.
\newblock {\em Phys. Lett.}, B392:77--84, 1997.

\bibitem[FQS18]{Fiset:2017auc}
Marc-Antoine Fiset, Callum Quigley, and Eirik~Eik Svanes.
\newblock {Marginal deformations of heterotic G$_{2}$ sigma models}.
\newblock {\em JHEP}, 02:052, 2018.

\bibitem[Fri80]{Friedan:1980jf}
D.~Friedan.
\newblock {Nonlinear Models in Two Epsilon Dimensions}.
\newblock {\em Phys. Rev. Lett.}, 45:1057, 1980.

\bibitem[GFRT15]{Garcia-Fernandez:2015hja}
Mario Garcia-Fernandez, Roberto Rubio, and Carl Tipler.
\newblock {Infinitesimal moduli for the Strominger system and Killing spinors
  in generalized geometry}.
\newblock 2015.

\bibitem[GFRT18]{Garcia-Fernandez:2018ypt}
Mario Garcia-Fernandez, Roberto Rubio, and Carl Tipler.
\newblock {Holomorphic string algebroids}.
\newblock 2018.

\bibitem[GGMT87]{Gates:1986ez}
S.~J. Gates, Jr., Marcus~T. Grisaru, L.~Mezincescu, and P.~K. Townsend.
\newblock {(1,0) SUPERGRAPHITY}.
\newblock {\em Nucl. Phys.}, B286:1--26, 1987.

\bibitem[GK04]{Gaberdiel:2004vx}
Matthias~R. Gaberdiel and Peter Kaste.
\newblock {Generalized discrete torsion and mirror symmetry for g(2)
  manifolds}.
\newblock {\em JHEP}, 08:001, 2004.

\bibitem[GKMW01]{Gauntlett:2001ur}
Jerome~P. Gauntlett, Nakwoo Kim, Dario Martelli, and Daniel Waldram.
\newblock {Five-branes wrapped on SLAG three cycles and related geometry}.
\newblock {\em JHEP}, 11:018, 2001.

\bibitem[GM88]{Grignani:1987gk}
G.~Grignani and M.~Mintchev.
\newblock {The Effect of Gauge and Lorentz Anomalies on the Beta Functions of
  Heterotic $\sigma$ Models}.
\newblock {\em Nucl. Phys.}, B302:330--348, 1988.

\bibitem[GMW04]{Gauntlett:2003cy}
Jerome~P. Gauntlett, Dario Martelli, and Daniel Waldram.
\newblock {Superstrings with intrinsic torsion}.
\newblock {\em Phys. Rev.}, D69:086002, 2004.

\bibitem[GN95]{Gunaydin:1995ku}
Murat Gunaydin and Hermann Nicolai.
\newblock {Seven-dimensional octonionic Yang-Mills instanton and its extension
  to an heterotic string soliton}.
\newblock {\em Phys. Lett.}, B351:169--172, 1995.
\newblock [Addendum: Phys. Lett.B376,329(1996)].

\bibitem[GS84]{Green:1984sg}
Michael~B. Green and John~H. Schwarz.
\newblock {Anomaly Cancellation in Supersymmetric D=10 Gauge Theory and
  Superstring Theory}.
\newblock {\em Phys. Lett.}, 149B:117--122, 1984.

\bibitem[GvdVZ86]{Grisaru:1986px}
Marcus~T. Grisaru, A.~E.~M. van~de Ven, and D.~Zanon.
\newblock {Four Loop beta Function for the N=1 and N=2 Supersymmetric Nonlinear
  Sigma Model in Two-Dimensions}.
\newblock {\em Phys. Lett.}, B173:423--428, 1986.

\bibitem[GW86]{Gross:1986iv}
David~J. Gross and Edward Witten.
\newblock {Superstring Modifications of Einstein's Equations}.
\newblock {\em Nucl. Phys.}, B277:1, 1986.

\bibitem[HHT87]{Henty:1987wc}
J.~C. Henty, C.~M. Hull, and P.~K. Townsend.
\newblock {World Sheet Supergravity Anomaly Cancellation for the Heterotic
  String in a Ten-dimensional Supergravity Background}.
\newblock {\em Phys. Lett.}, B185:73--78, 1987.

\bibitem[HKS88]{Hamada:1987ph}
Ken-ji Hamada, Jiro Kodaira, and Juichi Saito.
\newblock {HETEROTIC STRING IN BACKGROUND GAUGE FIELDS}.
\newblock {\em Nucl. Phys.}, B297:637--652, 1988.

\bibitem[HN12]{Harland:2011zs}
Derek Harland and Christoph Nolle.
\newblock {Instantons and Killing spinors}.
\newblock {\em JHEP}, 03:082, 2012.

\bibitem[Hon72]{Honerkamp:1971sh}
J.~Honerkamp.
\newblock {Chiral multiloops}.
\newblock {\em Nucl. Phys.}, B36:130--140, 1972.

\bibitem[HP88]{Howe:1988cj}
Paul~S. Howe and G.~Papadopoulos.
\newblock {Further Remarks on the Geometry of Two-dimensional Nonlinear
  $\sigma$ Models}.
\newblock {\em Class. Quant. Grav.}, 5:1647--1661, 1988.

\bibitem[HP91a]{Howe:1991vs}
Paul~S. Howe and G.~Papadopoulos.
\newblock {A Note on holonomy groups and sigma models}.
\newblock {\em Phys. Lett.}, B263:230--232, 1991.

\bibitem[HP91b]{Howe:1991im}
Paul~S. Howe and G.~Papadopoulos.
\newblock {W symmetries of a class of d = 2 N=1 supersymmetric sigma models}.
\newblock {\em Phys. Lett.}, B267:362--365, 1991.

\bibitem[HP93]{Howe:1991ic}
Paul~S. Howe and G.~Papadopoulos.
\newblock {Holonomy groups and W symmetries}.
\newblock {\em Commun. Math. Phys.}, 151:467--480, 1993.

\bibitem[HPS88]{Howe:1986vm}
Paul~S. Howe, G.~Papadopoulos, and K.~S. Stelle.
\newblock {The Background Field Method and the Nonlinear $\sigma$ Model}.
\newblock {\em Nucl. Phys.}, B296:26--48, 1988.

\bibitem[HPT93]{Hull:1993ct}
C.~M. Hull, G.~Papadopoulos, and P.~K. Townsend.
\newblock {Potentials for (p,0) and (1,1) supersymmetric sigma models with
  torsion}.
\newblock {\em Phys. Lett.}, B316:291--297, 1993.

\bibitem[HS06]{Howe:2006si}
P.~S. Howe and Vid Stojevic.
\newblock {On the symmetries of special holonomy sigma models}.
\newblock {\em JHEP}, 12:045, 2006.

\bibitem[HT86a]{Hull:1985rc}
C.~M. Hull and P.~K. Townsend.
\newblock {Finiteness and Conformal Invariance in Nonlinear $\sigma$ Models}.
\newblock {\em Nucl. Phys.}, B274:349--362, 1986.

\bibitem[HT86b]{Hull:1986xn}
C.~M. Hull and P.~K. Townsend.
\newblock {World Sheet Supersymmetry and Anomaly Cancellation in the Heterotic
  String}.
\newblock {\em Phys. Lett.}, B178:187--192, 1986.

\bibitem[HT87]{Hull:1987pc}
C.~M. Hull and P.~K. Townsend.
\newblock {The Two Loop Beta Function for $\sigma$ Models With Torsion}.
\newblock {\em Phys. Lett.}, B191:115--121, 1987.

\bibitem[HT88]{Hull:1987yi}
C.~M. Hull and P.~K. Townsend.
\newblock {String Effective Actions From $\sigma$ Model Conformal Anomalies}.
\newblock {\em Nucl. Phys.}, B301:197--223, 1988.

\bibitem[Hul86a]{Hull:1985dx}
C.~M. Hull.
\newblock {Anomalies, Ambiguities and Superstrings}.
\newblock {\em Phys. Lett.}, 167B:51--55, 1986.

\bibitem[Hul86b]{Hull:1986kz}
C.~M. Hull.
\newblock {Compactifications of the Heterotic Superstring}.
\newblock {\em Phys. Lett.}, B178:357--364, 1986.

\bibitem[Hul86c]{Hull:1986hn}
C.~M. Hull.
\newblock {LECTURES ON NONLINEAR SIGMA MODELS AND STRINGS}.
\newblock In {\em {NATO Advanced Research Workshop on Superfield Theories
  Vancouver, British Columbia, Canada, July 25-August 6, 1986}}, pages 77--168,
  1986.

\bibitem[Hul86d]{Hull:1985zy}
C.~M. Hull.
\newblock {$\sigma$ Model Beta Functions and String Compactifications}.
\newblock {\em Nucl. Phys.}, B267:266--276, 1986.

\bibitem[HW85]{Hull:1985jv}
C.~M. Hull and Edward Witten.
\newblock {Supersymmetric Sigma Models and the Heterotic String}.
\newblock {\em Phys. Lett.}, B160:398--402, 1985.
\newblock [,398(1985)].

\bibitem[II05]{Ivanov:2003nd}
Petar Ivanov and Stefan Ivanov.
\newblock {SU(3) instantons and G(2), spin(7) heterotic string solitons}.
\newblock {\em Commun. Math. Phys.}, 259:79--102, 2005.

\bibitem[Iva01]{Ivanov:2001ma}
Stefan Ivanov.
\newblock {Connection with torsion, parallel spinors and geometry of spin(7)
  manifolds}.
\newblock 2001.

\bibitem[Iva10]{Ivanov:2009rh}
Stefan Ivanov.
\newblock {Heterotic supersymmetry, anomaly cancellation and equations of
  motion}.
\newblock {\em Phys. Lett.}, B685:190--196, 2010.

\bibitem[JQ18]{Jardine:2018sft}
Ian~T. Jardine and Callum Quigley.
\newblock {Conformal invariance of (0, 2) sigma models on Calabi-Yau
  manifolds}.
\newblock {\em JHEP}, 03:090, 2018.

\bibitem[Kar05]{MR2164593}
Spiro Karigiannis.
\newblock Deformations of {$G_2$} and {${\rm Spin}(7)$} structures.
\newblock {\em Canad. J. Math.}, 57(5):1012--1055, 2005.

\bibitem[Lam96]{Lambert:1995hs}
N.~D. Lambert.
\newblock {Two loop renormalization of massive (p, q) supersymmetric sigma
  models}.
\newblock {\em Nucl. Phys.}, B469:68--92, 1996.

\bibitem[McO11]{McOrist:2010ae}
Jock McOrist.
\newblock {The Revival of (0,2) Linear Sigma Models}.
\newblock {\em Int. J. Mod. Phys.}, A26:1--41, 2011.

\bibitem[MM88]{Mavromatos:1988jj}
N.~E. Mavromatos and J.~L. Miramontes.
\newblock {Zamolodchikov's C Theorem and String Effective Actions}.
\newblock {\em Phys. Lett.}, B212:33--40, 1988.

\bibitem[MMS18]{Melnikov:2017yvz}
Ilarion~V. Melnikov, Ruben Minasian, and Savdeep Sethi.
\newblock {Spacetime supersymmetry in low-dimensional perturbative heterotic
  compactifications}.
\newblock {\em Fortsch. Phys.}, 66(5):1800027, 2018.

\bibitem[MN84]{Moore:1984dc}
Gregory~W. Moore and Philip~C. Nelson.
\newblock {Anomalies in Nonlinear $\sigma$ Models}.
\newblock {\em Phys. Rev. Lett.}, 53:1519, 1984.
\newblock [,1519(1984)].

\bibitem[MS11]{Melnikov:2011ez}
Ilarion~V. Melnikov and Eric Sharpe.
\newblock {On marginal deformations of (0,2) non-linear sigma models}.
\newblock {\em Phys. Lett.}, B705:529--534, 2011.

\bibitem[MSS12]{Melnikov:2012hk}
Ilarion Melnikov, Savdeep Sethi, and Eric Sharpe.
\newblock {Recent Developments in (0,2) Mirror Symmetry}.
\newblock {\em SIGMA}, 8:068, 2012.

\bibitem[MSV99]{Malikov:1998dw}
Fyodor Malikov, Vadim Schechtman, and Arkady Vaintrob.
\newblock {Chiral de Rham complex}.
\newblock {\em Commun. Math. Phys.}, 204:439--473, 1999.

\bibitem[NS86]{Nemeschansky:1986yx}
Dennis Nemeschansky and Ashoke Sen.
\newblock {Conformal Invariance of Supersymmetric $\sigma$ Models on Calabi-yau
  Manifolds}.
\newblock {\em Phys. Lett.}, B178:365--369, 1986.

\bibitem[Oda89]{Odake:1988bh}
Satoru Odake.
\newblock {Extension of $N=2$ Superconformal Algebra and Calabi-yau
  Compactification}.
\newblock {\em Mod. Phys. Lett.}, A4:557, 1989.

\bibitem[Pol88]{Polchinski:1987dy}
Joseph Polchinski.
\newblock {Scale and Conformal Invariance in Quantum Field Theory}.
\newblock {\em Nucl. Phys.}, B303:226--236, 1988.

\bibitem[Pol07]{Polchinski:1998rq}
J.~Polchinski.
\newblock {\em {String theory. Vol. 1: An introduction to the bosonic string}}.
\newblock Cambridge Monographs on Mathematical Physics. Cambridge University
  Press, 2007.

\bibitem[PT95]{Papadopoulos:1994kj}
G.~Papadopoulos and P.~K. Townsend.
\newblock {Solitons in supersymmetric sigma models with torsion}.
\newblock {\em Nucl. Phys.}, B444:245--264, 1995.

\bibitem[Sen85]{Sen:1985qt}
Ashoke Sen.
\newblock {Equations of Motion for the Heterotic String Theory from the
  Conformal Invariance of the Sigma Model}.
\newblock {\em Phys. Rev. Lett.}, 55:1846, 1985.

\bibitem[Sen86a]{Sen:1985tq}
Ashoke Sen.
\newblock {Local Gauge and Lorentz Invariance of the Heterotic String Theory}.
\newblock {\em Phys. Lett.}, 166B:300--304, 1986.

\bibitem[Sen86b]{Sen:1986nm}
Ashoke Sen.
\newblock {Superspace Analysis of Local Lorentz and Gauge Anomalies in the
  Heterotic String Theory}.
\newblock {\em Phys. Lett.}, B174:277--279, 1986.

\bibitem[Str86]{Strominger:1986uh}
Andrew Strominger.
\newblock {Superstrings with Torsion}.
\newblock {\em Nucl. Phys.}, B274:253, 1986.

\bibitem[SV95]{Shatashvili:1994zw}
Samson~L. Shatashvili and Cumrun Vafa.
\newblock {Superstrings and manifold of exceptional holonomy}.
\newblock {\em Selecta Math.}, 1:347, 1995.

\bibitem[Wen15]{Wendland:2015rla}
Katrin Wendland.
\newblock {K3 en route From Geometry to Conformal Field Theory}.
\newblock In {\em {8th Summer School on Geometric, Algebraic and Topological
  Methods for Quantum Field Theory Villa de Leyva, Colombia, July 15-17,
  2013}}, 2015.

\bibitem[Wit95]{Witten:1994tz}
Edward Witten.
\newblock {Sigma models and the ADHM construction of instantons}.
\newblock {\em J. Geom. Phys.}, 15:215--226, 1995.

\bibitem[Zam86]{Zamolodchikov:1986gt}
A.~B. Zamolodchikov.
\newblock {Irreversibility of the Flux of the Renormalization Group in a 2D
  Field Theory}.
\newblock {\em JETP Lett.}, 43:730--732, 1986.
\newblock [Pisma Zh. Eksp. Teor. Fiz.43,565(1986)].

\bibitem[Zum79]{Zumino:1979et}
B.~Zumino.
\newblock {Supersymmetry and Kahler Manifolds}.
\newblock {\em Phys. Lett.}, 87B:203, 1979.

\end{thebibliography}

\end{document}